\documentclass[12pt]{article}

\usepackage{soul}
\usepackage{lipsum}
\usepackage[many]{tcolorbox}
\usepackage{cancel}

\newtcolorbox{cross}{blank,breakable,parbox=false,
  overlay={\draw[red,line width=5pt] (interior.south west)--(interior.north east);
    \draw[red,line width=5pt] (interior.north west)--(interior.south east);}}

\usepackage{tikz}
\usetikzlibrary{decorations.markings}
\usepackage{graphicx}
\usepackage{caption}
\usepackage{subcaption}
\usepackage{amsmath}
\usepackage{amsthm}
\usepackage{amsfonts}
\usepackage{amssymb}
\usepackage{mathtools} 
\usepackage{bbm} 

\usepackage[T1]{fontenc}
\usepackage[utf8]{inputenc}
\usepackage[english]{babel}
\usepackage{csquotes}

\usepackage{longtable}
\usepackage{array}
\newcolumntype{C}[1]{>{\centering\arraybackslash}p{#1}}

\usepackage{tasks}

\usepackage{natbib}
\newtheorem{definition}{Definition}
\newtheorem{remark}{Remark}
\newtheorem{theorem}{Theorem}

\newcommand*{\xdash}[1][2.0em]{\rule[0.5ex]{#1}{0.55pt}}

\newcommand{\dvq}[2]{\frac{\text{d}^{#1}}{\text{d}#2^{#1}}}

\usepackage{scalerel,stackengine}
\stackMath
\newcommand\reallywidehat[1]{%
\savestack{\tmpbox}{\stretchto{%
  \scaleto{%
    \scalerel*[\widthof{\ensuremath{#1}}]{\kern-.6pt\bigwedge\kern-.6pt}%
    {\rule[-\textheight/2]{1ex}{\textheight}}
  }{\textheight}%
}{0.5ex}}%
\stackon[1pt]{#1}{\tmpbox}%
}

\newcommand{\NCF}{N_{F}}
\newcommand{\e}{\text{e}}
\newcommand{\dd}{\text{d}}

\allowdisplaybreaks

\title{
\textbf{The SINC way: A fast and accurate approach to Fourier pricing} \\
\author{
Fabio Baschetti\thanks{Department of Statistics, University of Bologna, Italy E-mail: fabio.baschetti@studio.unibo.it} 
\and
Giacomo Bormetti\thanks{Department of Mathematics, University of Bologna, Italy E-mail: giacomo.bormetti@unibo.it} 
\and
Silvia Romagnoli\thanks{Department of Statistics, University of Bologna, Italy E-mail: silvia.romagnoli@unibo.it} 
\and 
Pietro Rossi\thanks{Prometeia S.p.A., Bologna, Italy E-mail: pietro.rossi@prometeia.it}
\thanks{A previous version of this manuscript circulated with the title `Rough Heston: The SINC way'}
}
\date{\today}
}

\begin{document}

\maketitle

\begin{abstract}
The goal of this paper is to investigate the method outlined by one of us (PR) in Cherubini et al. (2009) to compute option prices. We name it the SINC approach. While the COS method by Fang and Osterlee (2009) leverages the Fourier-cosine expansion of truncated densities, the SINC approach builds on the Shannon Sampling Theorem revisited for functions with bounded support. We provide several results which were missing in the early derivation: i) a rigorous proof of the convergence of the SINC formula to the correct option price when the support grows and the number of Fourier frequencies increases; ii) ready to implement formulas for put, Cash-or-Nothing, and Asset-or-Nothing options;  iii) a systematic comparison with the COS formula for several log-price models; iv) a numerical challenge against alternative Fast Fourier specifications, such as Carr and Madan (1999) and Lewis (2000); v) an extensive pricing exercise under the rough Heston model of Jaisson and Rosenbaum (2015); vi) formulas to evaluate numerically the moments of a truncated density. The advantages of the SINC approach are numerous. When compared to benchmark methodologies, \emph{SINC} provides the most accurate and fast pricing computation. The method naturally lends itself to price all options in a smile concurrently by means of Fast Fourier techniques, boosting fast calibration. Pricing requires to resort only to odd moments in the Fourier space.\\

\end{abstract}

\section{Introduction}\label{sec:int}

The search of numerically efficient approaches to price options is the subject of intensive research. This fact comes with no surprise, since the ubiquitous presence and crucial role played by contingent claims in modern finance. It can be affirmed that, when the characteristic function (CF for short) of the log-price process is known in analytic or semi-analytic form, the current widely accepted solution to the pricing problem is the COS method by~\cite{doi:10.1137/080718061}. COS -- a short-name for Fourier-cosine expansion -- builds on the idea that it is computationally convenient to transform the expectation of the payoff with respect to the risk-neutral probability density function (PDF for short) into a linear combination of products of Fourier-cosine coefficients of the payoff and the density.
To achieve this goal, the price to pay is the approximation of true PDF by a truncated version with bounded support, but the trick eventually reveals to be the crucial step to obtain an excellent pricing formula. \\ 
\indent Our paper leverages the same idea of truncating the PDF, due to one of us (PR) and outlined in~\cite{cherubinibook}, but from a different perspective. It exploits a well-known result which applies to periodic functions with limited bandwidth, i.e. the Shannon Sampling Theorem. The formal symmetry between the forward and backward Fourier transform readily provides the intuition that Shannon's result can be adapted to functions with limited support in the direct space. 
As an interesting outcome of the application of the Sampling Theorem, one can express Plain Vanilla put and call prices, and digital option constituents, as a Fourier-sinc expansion. Given that the sinc function is the Fourier transform of the rectangular function, it is not surprising that it may play a crucial role in representing expectations with respect to truncated densities. The convolution between the sinc function -- which conveys the information related to the bounded support -- and the Fourier transform of the Heaviside step function -- which characterizes the point of discontinuity of the digital options -- lends itself to analytic simplification by means of the Modified Hilbert transform. As a result, the option price can be represented as a series expansion which only requires the CF computation of the log-price process for odd moments. We refer to this method as \emph{SINC approach}. As an important contribution, in this paper we prove in a rigorous way that the numerical error induced by the PDF truncation and by approximating a double infinite Fourier series by a finite sum can be made arbitrary small.\\ 
\indent It is worth mentioning that both COS and \emph{SINC} need to know the CF in order to be applied, hence this compulsory request singles out the range of applications we can deal with. The literature on stochastic models where it is natural to work in the Fourier space is huge and ever growing (see ~\cite{cherubinibook} for an overview of the topic). The successful application of Fourier analysis to price options was pioneered by~\cite{chen1992pricing, heston1993closed,bates1996jumps, bakshi1997alternative,scott1997pricing}. The publication of~\cite{duffie2000transform} definitely celebrated the role of the transform analysis in dynamic asset pricing models when the state vector follows an affine jump-diffusion. The papers by~\cite{carr1999option} and~\cite{lewisbook,lewis2001simple} contributed in a significant way to this stream of research in quantitative finance. In the former, the authors introduced a simple analytical expression for the Fourier transform of the option value, which allows to exploit the considerable computational power of the Fast Fourier Transform (FFT) in the inversion stage. The introduction of FFT techniques boosted the way to real-time calibration, pricing, and hedging. In the latter contributions, ~\cite{lewisbook,lewis2001simple} detailed a representation of the option price in terms of the CF which is rooted on a clever extension of the Fourier transform in the complex domain. His approach is naturally prone to the application of FFT, too. It is often preferred to the ~\cite{carr1999option} approach, which requires the introduction of an auxiliary damping parameter. \emph{SINC} is naturally suited for the computation by means of FFT. Then, not only \emph{SINC} is based on a parsimonious representation of the option payoff, which requires to sample the CF at optimal points, but it expresses the payoff as a transform where the log-moneyness is the conjugate variable in the direct space. As a consequence, all option prices in a smile can be computed concurrently with $O(N\log_2 N)$ complexity, where $N$ is the number of sample points in Fourier space, enhancing the computational advantage of \emph{SINC} with respect to COS.  

 The stream of research inspired by the general framework introduced in~\cite{duffie2000transform} is vast. It ranges from models to equity and exchange rate option pricing, to interest rate derivative pricing, credit risk, and systemic risk modeling. Following~\citep{doi:10.1137/080718061}, first we test the performances of \emph{SINC} on commonly used stochastic models for the equity log-price process, i.e. the Geometric Brownian motion (GBM), the Heston model~\citep{heston1993closed}, and the CGMY model by~\citep{carr2002fine}. Then, we focus on a restricted but stimulating and flourishing field, the modeling of financial volatility for pricing purposes~\footnote{Monte Carlo methods represent an alternative approach to pricing under rough volatility. We do not consider it here, because it is quite aside from our main message. We refer the interested reader to~\citep{bennedsen2017hybrid,mccrickerd2018turbocharging,bayer2020hierarchical} for recent developments.}. The main reason of the interest in volatility modeling is that, recently, the celebrated Heston model has been revisited in several respects. 
\cite{jaisson2015limit} showed that the Hawkes-based~\citep{hawkes1971point,hawkes1971spectra} market microstructure model of~\cite{bacry2013modelling} under nearly-unstable conditions converges in law to the Heston model. \cite{jaisson2016rough} also proved that the microstructure model with an hyperbolic kernel by~\citep{hardiman2013critical,bacry2016estimation} converges to an integrated fractional diffusion. The limiting process is very irregular, with a derivative behaving as a fractional Brownian motion with Hurst exponent smaller than 0.5 and close to zero. For this reason, it is dubbed rough Heston (rHeston for short). \cite{gatheral2018volatility} demonstrated for a wide range of assets that the historical volatility is rougher than a Brownian motion, and that the empirical moment of order $q$ of the log-volatility increments are consistent with a scaling with Hurst exponent of order 0.1. Similar findings are reported in~\cite{bennedsen2016decoupling} under the historical measure, while \cite{livieri2018rough} investigated the rough behavior of the implied volatility. Finally, contrary to classical volatility models, the rough ones (and so also the rHeston model) are able to reproduce the explosive behavior of the implied at-the-money (ATM) skew observed empirically when the option maturity goes to zero~\citep{bayer2016pricing,Fukasawa2011}. Remarkably, \cite{el2019characteristic} derived a semiclosed formula for the CF of rHeston model. The formula is not fully explicit but given in terms of the solution of a fractional Riccati equation; the equation admits a unique continuous solution, whose closed form expression is unknown. To avoid the computational burden arising from the numerical solution of the fractional Riccati equation, in this paper we resort to the Padè approximant of the solution already discussed in~\cite{gatheral2019rational}. As shown by the authors, the rational approximation provides a very accurate description of the solution, especially for low values of the Hurst exponent $H$. In empirical investigations, both under the pricing and the historical measures, $H$ is found to be of order 0.05-0.1, thus motivating the use of the rational approximation. An alternative approach is provided by the Adams scheme~\citep{diethelm2004detailed}, possibly combined with a power series expansion and Richardson-Romberg extrapolation~\citep{callegaro2020fast}.\\

\indent As a second main contribution of our paper, we challenge \emph{SINC} against COS and \emph{FFT-SINC} against ~\cite{carr1999option}  and ~\cite{lewisbook,lewis2001simple}  approaches computed via FFT. Through extensive pricing under the forward variance specification, we assess the superiority in pricing accuracy of \emph{SINC} with respect to competitors. The comparison is performed keeping the same number $\NCF$ of points sampled in the Fourier space equal for all methodologies. We believe this is the fairest way to claim the relative performance of the different algorithms, since the number of times the CF needs to be computed in rHeston represents the most time consuming step in pricing. Under this specification, when \emph{SINC} is challenged against COS, the superiority of the former is apparent. When the full power of FFT is exploited, the numerical complexity reduction of \emph{SINC} vs COS is sizable and dramatic, making \emph{SINC} our preferred approach. As a matter of fact when dealing with the rHeston model, the main computational burden comes from the solution of the fractional Riccati equation needed to get the CF. This part greatly outweights the cost of pricing even a highly populated smile and the burden of using FT is twice as big as that of FFT in our exercise. Very much different is the case where the CF is known analitically; in that case the advantage of having a natural FFT formulation would be very large.\\

\indent Last, but not least, as a side result of \emph{SINC} approach, we detail in the Appendix a novel analytical methodology to approximate the moments of a random variable starting from the CF.\\

The remainder of the paper is organized as follows. In Section~\ref{sec:SINC} we discuss the \emph{SINC} formula and in Section~\ref{section:ErrAn} we characterize the numerical error. Sections ~\ref{section:section 03} and ~\ref{section:section 05}  present the numerical results from the pricing  exercise by means of the \emph{SINC} and FFT-\emph{SINC} specifications, respectively. Section~\ref{sec:concl} draws the most relevant conclusions. The Appendix provides technical details.

\section{\emph{SINC} at a glance}\label{sec:SINC}

The \emph{SINC} approach to price options is rooted on the following definition of a Fourier pair
\begin{align*}
& g(x) = \bar{\mathcal{F}}[\hat{g}(\kappa)] = \int_\mathbb{R} \e^{-i 2\pi x \kappa} \hat{g}(\kappa) \dd \kappa,\\ 
& \hat{g}(\kappa) = \mathcal{F}[g(x)] = \int_\mathbb{R} \e^{+i 2\pi x \kappa} g(x) \dd x, 
\end{align*} 
where $\bar{\mathcal{F}}$ and $\mathcal{F}$ stand for the forward Fourier operator and the inverse Fourier operator, respectively and $g(.)$ is integrable.
Under the assumption of null interest rate and dividend yield, i.e. $r=0$ and $q=0$, it exploits the following decomposition of a Plain Vanilla (PV hereafter) put into Cash or Nothing (CoN hereafter) plus Asset or Nothing (AoN) options, i.e.
\begin{align}
\mathbb{E}[(K - S_T)^+] & = K \mathbb{E}[\mathbbm{1}_{\{s_T<k\}}] - S_0 \mathbb{E}[\e^{s_T} \mathbbm{1}_{\{s_T<k\}}], \hspace{7mm} s_T = \log \bigg( \frac{S_T}{S_0} \bigg), \quad k = \log \bigg( \frac{K}{S_0} \bigg) \label{put_price}
\end{align}
with $S_T$ and $K$ denoting the underlying spot at time $T$ and the exercise price, respectively~\footnote{The general formula for non zero interest rate and dividend yield is readily recovered by setting $s_T = \log (S_T/S_0) - (r-q)T$ and $k = \log (K/S_0) - (r-q)T$ and reads 
$\text{Put}(t=0,S_0)=\e^{-qT}S_0\mathbb{E}[(\e^k-\e^{s_T})^+]\,.$}.\\

\noindent We note as $\theta(x)$ the Heaviside step function and recognize that contour integration yields
\begin{align*}
\theta(x) = \bar{\mathcal{F}}[\delta^-(\kappa)] = \int \e^{-i 2\pi \kappa x} \delta^-(\kappa) \dd \kappa,
\end{align*}
where $\delta^- (\kappa) = \frac{i}{2\pi} \frac{1}{\kappa+i\varepsilon}$. In the Appendix (Section~\ref{appendix-inv-fft-theta}), we recall the derivation of the previous result and clarify the role played by $\varepsilon$.\\

\noindent Therefore, if we write each of the expectations on the rhs of Equation~(\ref{put_price}) in terms of the PDF of the log-return $s_T$, $f(s_T)$, and the payoff of the option, we have that
\begin{align}
\mathbb{E}[\mathbbm{1}_{\{s_T<k\}}] & = \int f(s_T) \theta(k-s_T) \dd s_T = \bar{\mathcal{F}} \big[ \mathcal{F}[f(k)] \mathcal{F}[\theta(k)] \big] = \bar{\mathcal{F}}[\hat{f}(\kappa) \delta^-(\kappa)] \nonumber \\
& = \frac{i}{2\pi} \int \e^{-i 2\pi k \kappa} \hat{f}(\kappa) \frac{1}{\kappa+i\varepsilon} \dd \kappa\,, \label{CoN_01} 
&&
\end{align}
and
\begin{align}
\mathbb{E}[\e^{s_T} \mathbbm{1}_{\{s_T<k\}}] & = \int \e^{s_T} f(s_T) \theta(k-s_T) \dd s_T = \frac{i}{2\pi} \int \e^{-i 2\pi k \kappa} \hat{f} \bigg( \kappa - \frac{i}{2\pi} \bigg) \frac{1}{\kappa+i\varepsilon} \dd \kappa 
\end{align}
by simple means of the convolution theorem and the definition of a Fourier transform (FT for short). \\

\noindent Observe that a change of measure is implicit in the expectation defining the AoN put, which requires that $\mathbb{E}[\e^{s_T}]=1$.\\

\noindent For any given $\eta>0$, we can find $X_l$ and $X_h$ for which  
\begin{align*}
\bigg| 1 - \int_{X_l}^{X_h} f(s_T) \dd s_T \bigg| < \eta,
\end{align*}
and the Shannon Sampling Theorem \citep{shannon1949communication} guarantees that the Fourier transform of the truncated function $f(s_T)\mathbbm{1}_{X_l\leq s_T \leq X_h}$ can be fully recovered given a discrete (countable) set of points. Indeed, in the Appendix (Section \ref{appendix:Appendix A}), we show that
\begin{align}
\e^{-i 2\pi k \kappa} \reallywidehat{f\mathbbm{1}_{\{X_l\leq s_T \leq X_h\}}}(\kappa) = \sum_{n=-\infty}^\infty \e^{-i 2\pi k \kappa_n} \reallywidehat{f\mathbbm{1}_{\{X_l\leq s_T \leq X_h\}}}(\kappa_n) \text{sinc}[2\pi X_c (\kappa_n - \kappa)], \label{Shannon}
\end{align}
where $\kappa_n = n/(2X_c), \ X_c = (X_h - X_l)/2$, and the sinc function is defined in the usual way as $\sin(x)/x$ (continuous at zero). \\

\noindent In other terms, the idea is that one can truncate the integration range in such a way that the contribution from the tails of the PDF is arbitrarily small, and getting rid of it provides an upper bound for the approximation error induced on the option price. As we are working with Fourier transforms, it is convenient to think of the length of the truncation range $2X_c$ as the periodicity of the bounded density. Then, we suggest that $X_l$ and $X_h$ are selected  according to the following constraints 
\begin{align}
\int_{-\infty}^{X_l} f(s_T) \dd s_T < 10^{-10}, \qquad
\int_{-\infty}^{X_h} f(s_T) \dd s_T > 1-10^{-10}.\label{eq:cutting}
\end{align}
We will not allow for asymmetric intervals in our numerical sections and impose $X_h=-X_l=X_c$, for the sake of simplicity. Operationally, we compute the previous inequalities by means of Equation~(\ref{CoN_04}) and by choosing some safely large candidates for $X_l$ and $X_h$, then define $X_c = 4\max(|X_l|,|X_h|)$ and iterate until the difference between the new candidate value for $X_c$ and the old one is less than 30\%. \\

\noindent The need for truncating the density of the asset log-price is nothing new in the context of Fourier methods and it exactly motivates COS formulas by~\cite{doi:10.1137/080718061}. They in fact come up with one handy rule for determining the bounds of the PDF which reads as follows 
\begin{align}
[X_l,X_h] = \bigg[ c_1-L\sqrt{c_2+\sqrt{c_4}}, c_1+L\sqrt{c_2+\sqrt{c_4}} \bigg]. \label{trunc_rule}
\end{align}
Here $c_n$ tags the $n$-th cumulant of $s_T$ and $L$ is an arbitrary constant that mostly depends on the particular model one is considering. If this has the merit of being particularly simple, it still suffers from two problems: (i) it does not provide any clue as to the magnitude of the error associated with the truncation, and (ii) it requires knowledge of quantities which are not always given in closed form (just think about the rHeston model, for example). The second limitation may be overcome by numerical evaluation of the moments of the distribution -- we provide original formulas for doing so in the Appendix (Section \ref{appendix:Appendix D})-- but the former is a very strong reason for preferring the cutting strategy we have described in the previous paragraph. \\

\noindent We will provide an explicit formula for the numerical evaluation of the cumulative distribution function (CDF) in~(\ref{eq:cutting}) by the end of this section, and address to the numerical experiments for an assessment of its performance. It eventually turns out that the \emph{SINC} is an excellent way to compute distribution functions, which fact makes our procedure for the bounds of the PDF particularly cheap. This may clearly be extended to the COS, but the evaluation of the CDF would be much more costly. In this regard, one may also observe that the periodicity of the PDF in the COS method is actually $4X_c$ but the support coincides with the \emph{SINC}. \\

\noindent Now, we are in the position to recover both CoN and AoN put prices. Nevertheless, we only keep track of the CoN put for making things concise.\footnote{ The derivation of the AoN put price is perfectly equivalent to the CoN one. We decide to skip it because going through each steps would not add anything new.} As we have seen, bounding the PDF allows for an application of the Sampling Theorem: we plug Shannon's representation (\ref{Shannon}) into the CoN Equation~(\ref{CoN_01}), straightforwardly write
\begin{align}
\mathbb{E}[\mathbbm{1}_{\{s_T<k\}}] & \simeq \mathbb{E}[\mathbbm{1}_{\{s_T<k\}} \mathbbm{1}_{\{X_l\leq s_T \leq X_h\}}] \nonumber \\ & = \frac{i}{2\pi} \sum_{n=-\infty}^{\infty} \e^{-i 2\pi k \kappa_n} \reallywidehat{f\mathbbm{1}_{\{X_l\leq s_T \leq X_h\}}}(\kappa_n) \int \frac{\text{sinc}[2\pi X_c (\kappa_n - \kappa)]}{\kappa + i\varepsilon} \dd \kappa \label{CoN_02}
\end{align}
and finally recognize the inner integral in the sinc as a Modified Hilbert transform $\mathcal{H^-}$.

\begin{definition}\label{modHilb}
The Modified Hilbert transform $\mathcal{H^-}$ of a given function $g$ is the result of a convolution of the distribution $\delta^-(x)$ with the function itself. This formally translates as:
\begin{align*}
\mathcal{H}^-[g(y)] = \int g(y-x) \delta^-(x) \dd x = \frac{i}{2\pi} \int \frac{g(y-x)}{x+i\varepsilon} \dd x.
\end{align*}
\end{definition}

\noindent In particular, the Appendix (Section \ref{appendix:Appendix B}) proves that
\begin{align}
\int \frac{\text{sinc}[2\pi X_c (\kappa_n - \kappa)]}{\kappa + i\varepsilon} \dd \kappa = \frac{2\pi}{i} \mathcal{H^-}[\text{sinc}(2\pi X_c \kappa_n)] = \frac{1}{2X_c \kappa_n}(1-\e^{i 2\pi X_c \kappa_n}), \label{sinc_Hilbert}
\end{align}
which is sufficient to specialize the CoN put as
\begin{align*}
\mathbb{E}[\mathbbm{1}_{\{s_T<k\}}] \simeq \frac{i}{2\pi} \sum_{n=-\infty}^{\infty} \e^{-i 2\pi k \kappa_n} \reallywidehat{f\mathbbm{1}_{\{X_l\leq s_T \leq X_h\}}}(\kappa_n) \bigg[ -i\pi \mathbbm{1}_{n=0} + \frac{1-(-1)^n}{n} \mathbbm{1}_{n\neq 0} \bigg]. 
\end{align*}
An additional approximation is introduced when truncating this last infinite sum to a finite (possibly low) number of terms and the price of the CoN option is written accordingly as
\begin{align}
& \mathbb{E}[\mathbbm{1}_{\{s_T<k\}}] \simeq \frac{i}{2\pi} \sum_{n=-N/2}^{N/2} \e^{-i2\pi k \kappa_n} \reallywidehat{f\mathbbm{1}_{\{X_l\leq s_T \leq X_h\}}}(\kappa_n) \bigg[ -i\pi \mathbbm{1}_{n=0} + \frac{1-(-1)^n}{n} \mathbbm{1}_{n\neq 0} \bigg]\,. \nonumber
\end{align}
Then the final formula follows replacing $\reallywidehat{f\mathbbm{1}_{\{X_l\leq s_T \leq X_h\}}}(\kappa_n)$ with $\hat{f}(\kappa_n)$ and recognizing that only the odd moments in the Fourier space are relevant for the computation
\begin{align}
& \mathbb{E}[\mathbbm{1}_{\{s_T<k\}}] \simeq \frac{1}{2} + \frac{2}{\pi} \sum_{n=1}^{N/4} \frac{1}{2n-1} \bigg[ \sin(2\pi k \kappa_{2n-1}) \Re\big[\hat{f}(\kappa_{2n-1})\big] - \cos(2\pi k \kappa_{2n-1}) \Im \big[\hat{f}(\kappa_{2n-1})\big] \bigg]\,. \label{CoN_04}
\end{align}
Here $\Re$ and $\Im$ denote the real and imaginary parts, respectively. We show the validity of this final formula in the Appendix (Section \ref{appendix:Appendix C}), and claim that the AoN option is priced in a very similar way, except that the CF needs to be evaluated for a complex argument, i.e. 
\begin{align}
& \mathbb{E}[\e^{s_T}\mathbbm{1}_{\{s_T<k\}}] \simeq \frac{i}{2\pi} \sum_{n=-N/2}^{N/2} \e^{-i2\pi k \kappa_n} \hat{f}(\kappa_n-\frac{i}{2\pi}) \bigg[ -i\pi \mathbbm{1}_{n=0} + \frac{1-(-1)^n}{n} \mathbbm{1}_{n\neq 0}  \bigg] \nonumber \\
& = \frac{1}{2} + \frac{2}{\pi} \sum_{n=1}^{N/4} \frac{1}{2n-1} \bigg[ \sin(2\pi k \kappa_{2n-1}) \Re\big[\hat{f}(\kappa_{2n-1}-\frac{i}{2\pi})\big] \nonumber \\ 
&\hspace{78.5mm} - \cos(2\pi k \kappa_{2n-1}) \Im \big[\hat{f}(\kappa_{2n-1}-\frac{i}{2\pi})\big] \bigg]. \label{AoN_04}
\end{align}

\begin{remark}\label{pos_odd_freq}
Out of the $N+1$ terms that we included in the expansions, only $N/4$ survive. They correspond to the positive odd frequencies.
\end{remark}

\noindent While our cutting procedure provides us with explicit bounds on the PDF truncation error, we clearly need to control the impact of early termination of the infinite Fourier series and the usage of the CF for the complete density in place of the reduced one. We consequently decompose the overall error as the sum of the three components described in Section \ref{section:ErrAn} and study their behavior in the Appendix (Section \ref{appendix:error}): each of them is bounded and the error shrunk under suitable choice of $N$ and $X_c$, thus ensuring convergence of the $\emph{SINC}$ formulas to the true option price.  \\ 

\noindent The puzzle is finally complete when we combine digital options to compute PV put prices: 
\begin{theorem} Let $\hat{f}$ denote the CF of the asset log-return $s_T=\log(S_T/S_0)$ and take $k=\log(K/S_0)$ the log-moneyness of the option. Then Equations (\ref{put_price}), (\ref{CoN_04}) and (\ref{AoN_04}) justify the following writing of a PV put price
\begin{align}
\mathbb{E}[(K - S_T)^+] & \simeq \frac{1}{2}(K-S_0) \nonumber \\
& + \frac{2}{\pi} \sum_{n=1}^{N/4} \frac{1}{2n-1} \bigg[ \sin(2\pi k \kappa_{2n-1}) \Re\big[K\hat{f}(\kappa_{2n-1})-S_0\hat{f}(\kappa_{2n-1}-\frac{i}{2\pi})\big]\nonumber\\ 
& - \cos(2\pi k \kappa_{2n-1}) \Im \big[K\hat{f}(\kappa_{2n-1})-S_0\hat{f}(\kappa_{2n-1}-\frac{i}{2\pi})\big] \bigg]\,\label{eq:put}
\end{align}
where $\kappa_n = \frac{n}{X_h-X_l}$ and the interval $[X_l,X_h]$ is chosen so as to make the contribution from the tails of the PDF negligible.
\end{theorem}

\noindent To ease the interpretation of the results in the numerical sections and the comparison among different benchmark methodologies, we introduce the notation $\NCF$ to refer to the number of times the CF needs to be evaluated to compute the option price. For instance, to price a CoN put, it is sufficient to sample the CF $\NCF=N/4$ times at points $\kappa_{2n-1}$ ($N/4$ times at shifted points $\kappa_{2n-1}-i/(2\pi)$ for the AoN put) and to weight them with a suitable imaginary phase and the inverse of the integer odd numbers. The price of the PV put is readily recovered from AoN and CoN, thus by means of $\NCF=N/2$ valuations of the CF.  In the next sections, we are going to support the computational effectiveness of the \emph{SINC} formulas, by challenging them against the COS ones and showing how the \emph{SINC} approach can be readily adapted to the FFT framework.    

\subsection{The FFT form of \emph{SINC}} \label{FFT-SINC}

\noindent One merit of \emph{SINC} is that it is readily adapted to the stiff structure of the FFT algorithm. The computational speed of the Fast Fourier Transform is crucial for any concrete application within the calibration process and the extension comes with almost no effort in our setting. \\

\noindent We work under the assumption to price a discrete grid of strikes $k_m = m \frac{2X_c}{N}, -N/2 \leq m < N/2 $
and to fit the remaining points, when needed, by linear interpolation from bucket to bucket. \\ 

\noindent Digital put prices at the aforementioned vector of strikes are now calculated as follows
\begin{align}
\mathbb{E}[\e^{as_T} \mathbbm{1}_{\{s_T<k_m\}}] & \simeq \frac{i}{2\pi} \sum_{n=-N/2}^{N/2} \e^{-i 2\pi k_m \kappa_n} \hat{f}\bigg(\kappa_n-a\frac{i}{2\pi}\bigg) \bigg[ -i\pi \mathbbm{1}_{n=0} + \frac{1-(-1)^n}{n} \mathbbm{1}_{n\neq 0} \bigg] \nonumber\\
& = \frac{i}{2\pi} \sum_{n=0}^{N-1} \e^{-i \frac{2\pi}{N} mn} q_n  \label{eq:FFT_digital}
\end{align}
where
\begin{align}
q_n =
\begin{cases}
\frac{\pi}{i} & n = 0 \\
\hat{f}(\kappa_n-a\frac{i}{2\pi}) \frac{1-(-1)^n}{n} & n \in [1, \frac{N}{2}) \\
0 & n = \frac{N}{2} \\
\hat{f}(\kappa_{n-N}-a\frac{i}{2\pi}) \frac{1-(-1)^{n-N}}{n-N} & n \in (\frac{N}{2}, N-1]
\end{cases} \label{dig_FFT}
\end{align}
and $a$ takes value 0 or 1 for CoN and AoN options, respectively. Equation~(\ref{eq:FFT_digital}), taken together with the definition of $q_n$ in~(\ref{dig_FFT}), expresses the \emph{SINC} formulas in a form which can be readily computed by means of FFT. The formula for the PV put follows as before.
\begin{remark}
In spite of the fact that the index $n$ runs from $0$ to $N-1$, a closer inspection reveals that the computation of $q_n$ only requires the evaluation of the CF at $N/4$ different frequencies. Indeed, all $q_n$ for even $n$ are identically zero. 
\end{remark}

\noindent The described procedure generates prices for CoN and AoN digitals indexed by the strikes $n (2X_c/N)$. To recover the price for different strikes (not belonging to the grid) we perform a linear interpolation. The interpolation error can be reduced by increasing the number of terms in the expansion or resorting to the fractional FFT (frFFT) framework. 

\section{Error Analysis}\label{section:ErrAn}

As already mentioned, and similarly to the COS method of \cite{doi:10.1137/080718061}, there are three sources of error affecting the SINC formula: the approximation of the true PDF with a truncated density, the replacement of a double infinite sum with a finite sum, and the substitution of the Fourier coefficients for the truncated density with the Fourier transform of the true PDF valued at discrete points.
To characterize in a quantitative way the three error components, we proceed as follows.\\
The error associated to our approach can be written as~\footnote{As done before for the pricing formula, we detail the case of CoN put options. Similar results for the AoN puts can be readily derived.}
\begin{eqnarray}
    \epsilon & = &  \int f(s_T) \theta( k - s_T )\, \dd s_T - \frac{1}{2} -\frac{i}{2\pi X_c} \sum_{n = -N/4}^{+N/4} \e^{-i2\pi k\kappa_{2n-1}}\; \frac{\hat{f}(\kappa_{2n-1})}{\kappa_{2n-1}} \nonumber 
    \\ & = &  \int f(s_T) \theta( k - s_T )\, \dd s_T -  \int_{-X_c}^{X_c} f(s_T) \theta( k - s_T )\, \dd s_T  \nonumber
\\ & + &   \int_{-X_c}^{X_c} f(s_T) \theta( k - s_T )\, \dd s_T  
  - \frac{1}{2} -\frac{i}{2\pi X_c} \sum_{n = -N/4}^{+N/4} \e^{-i2\pi k\kappa_{2n-1}}\; \frac{\hat{f}(\kappa_{2n-1})}{\kappa_{2n-1}}\,. \nonumber 
\end{eqnarray}
Exploiting the fact that
\begin{eqnarray}
\int_{-X_c}^{X_c} f(s_T) \theta( k - s_T )\, \dd s_T  & = &  \frac{1}{2} + \frac{i}{2\pi X_c}\sum_{-\infty}^{+\infty} \e^{-i2\pi k\kappa_{2n-1}}\; \frac{\reallywidehat{f\mathbbm{1}_{\{-X_c\leq s_T \leq X_c\}}}(\kappa_{2n-1})}{\kappa_{2n-1}}\,, \nonumber
\end{eqnarray}
we can write 
\begin{eqnarray}
    \epsilon & = &  \int f(s_T) \theta( k - s_T )\, \dd s_T -  \int_{-X_c}^{X_c} f(s_T) \theta( k - s_T )\, \dd s_T  \nonumber
\\ & + &   \frac{i}{2\pi X_c}\sum_{-\infty}^{+\infty} \e^{-i2\pi k\kappa_{2n-1}}\; \frac{\reallywidehat{f\mathbbm{1}_{\{-X_c\leq s_T \leq X_c\}}}(\kappa_{2n-1})}{\kappa_{2n-1}} 
   -\frac{i}{2\pi X_c} \sum_{n = -N/4}^{+N/4} \e^{-i2\pi k\kappa_{2n-1}}\; \frac{\hat{f}(\kappa_{2n-1})}{\kappa_{2n-1}} \nonumber 
    \\ & = &  \int f(s_T) \theta( k - s_T )\, \dd s_T -  \int_{-X_c}^{X_c} f(s_T) \theta( k - s_T )\, \dd s_T  \nonumber
    \\   & + & \frac{i}{2\pi X_c} \sum_{|n| >N/4} \e^{-i2\pi k\kappa_{2n-1}}\; \frac{\reallywidehat{f\mathbbm{1}_{\{-X_c\leq s_T \leq X_c\}}}(\kappa_{2n-1})}{\kappa_{2n-1}} \nonumber 
\\ & + &   \frac{i}{2\pi X_c}\sum_{-N/4}^{+N/4} \e^{-i2\pi k\kappa_{2n-1}}\; \frac{\reallywidehat{f\mathbbm{1}_{\{-X_c\leq s_T \leq X_c\}}}(\kappa_{2n-1}) - \hat{f}(\kappa_{2n-1})}{\kappa_{2n-1}}\,.  \label{eq:error}
\end{eqnarray}
The PDF truncation error reads
$$\epsilon_1 \doteq \int f(s_T) \theta( k - s_T )\, \dd s_T -  \int_{-X_c}^{X_c} f(s_T) \theta( k - s_T )\, \dd s_T\,,$$
where we introduce the same notation, $\epsilon_1$, used in~\cite{doi:10.1137/080718061}. The second and last components of the error in Equation~(\ref{eq:error}), that we refer to with $\epsilon_2$ and $\epsilon_3$ to conform with the notation in~\cite{doi:10.1137/080718061}, are the error contributions due to the truncation of a double infinite Fourier series and the replacement of the Fourier coefficients of the truncated PDF with the Fourier transform of the true PDF, respectively. \\

\noindent Such a decomposition of the overall error is the starting point to prove that the \emph{SINC} price converges to the true option price: technical reasons and assumptions essential for the proof are given in the Appendix (Section \ref{appendix:error}), where we bound the magnitude for each of the components in Equation~(\ref{eq:error}) and conclude that the error can be made arbitrarily small by increasing the number of Fourier modes $N$ and the truncation range $[-X_c,X_c]$. 

\section{\emph{SINC} at work}\label{section:section 03}

In this section, we perform numerical tests to assess the accuracy of the \emph{SINC} approach. We price PV puts and their digital components separately and span over various maturities and moneynesses for both standard models (GBM, Heston, and CGMY) and the rHeston model of \cite{eleuch2018}. The idea is to compare \emph{SINC} and COS methods along the directions of precision and convergence speed. The COS is the most used method within the class of FT-based techniques. This is due to the acknowledged performance both in speed and accuracy. It is therefore natural to use it as reference. \\

\noindent The results we will produce show that the \emph{SINC} is often better than COS when computing call and put option prices. Always orders of magnitude better when dealing with digital options. Furthermore, \emph{SINC} enjoys the non negligible advantage to be tailor made for the FFT, while the COS, as we know, does not have a straightforward transition. The consequences of this will be more extensively discussed in the following section, where we also compare the \emph{SINC} with standard FFT methods.\\

\noindent While option prices in the Black-Scholes model have a closed-form solution to be used as a benchmark we need to produce one benchmark for the other models. We decided to use the average between high precision ($\NCF=2^{20}$) \emph{SINC} and COS prices and to chop them if the two happen to have more than 10 decimal digits in common.

\subsection{Geometric Brownian Motion (GBM)}

We begin with the simple example of a GBM for the price process. Selected parameters are the same as in \cite{doi:10.1137/080718061}
\begin{align*}
S_0 = 1 \quad r = 0.1 \quad q = 0 \quad T = 0.1 \quad \sigma = 0.25
\end{align*}
and nine different strikes are considered, $K=0.80:0.10:1.40$. This will be common throughout the section. \\

\noindent Table \ref{GBM_conv} reports relative errors with respect to the Black-Scholes price for both \emph{SINC} and COS at different values of $\NCF$. PV (lhs) and CoN (rhs) put options are considered, here. \\

\begin{table}[h!]
\tiny \centering
\caption{Relative errors over PV and CoN put options for \emph{SINC} and COS at different values of $\NCF$ in the Black-Scholes model. Stars ($\star$) mean that the price fully conforms with the benchmark (up to the number of digits of the benchmark itself), straight lines denote relative errors that are larger than 100\%. [$T=0.1$, $X_c= 2.0105$]}
\begin{tabular}{| l | c | c c c c c c |c| c c c c c c |c|}
 \hline
 & & \multicolumn{7}{c|}{PV put} & \multicolumn{7}{c|}{CoN put}\\
 \hline
 & & \multicolumn{6}{c|}{$\NCF$} & & \multicolumn{6}{c|}{$\NCF$} & \\
 \hline
 & K & 20 & 40 & 60 & 80 & 100 & 120 & benchmark & 20 & 40 & 60 & 80 & 100 & 120 & benchmark \\
 \hline
 SINC & 0.60 & \xdash & \xdash  & 5e-03 & 5e-04 & 5e-04 & 5e-04 & 1.974722e-13 & \xdash & 3e-06 & 3e-06 & 3e-06 & 3e-06 & 3e-06 & 1.726736e-11 \\
 COS  &      & \xdash & \xdash & \xdash & \xdash & \xdash & 2e-04 & & \xdash & \xdash & \xdash & \xdash & 4e-01 & 1e-04 & \\
\hline
 SINC & 0.70 & \xdash & 4e-01 & 1e-07 & 1e-08 & 2e-09 & 2e-09 & 2.301833e-08 & 2e-02 & 4e-11 & 1e-11 & 1e-11 & 1e-11 & 1e-11 & 1.473992e-06 \\
 COS  &      & \xdash & \xdash & \xdash & 1e-02 & 2e-05 & 4e-09 & & \xdash & \xdash & \xdash & 2e-02 & 1e-05 & 3e-10 & \\
\hline
 SINC & 0.80 & \xdash & 3e-04 & 9e-11 & 2e-12 & 2e-12 & 2e-12 & 3.2130086e-05 & 2e-05 & 2e-15 & 6e-14 & 6e-14 & 6e-14 & 6e-14 & 0.0014334833  \\
 COS  &      & \xdash & 2e-01 & 2e-03 & 1e-05 & 1e-08 & 3e-12 & & \xdash & 8e-01 & 9e-03 & 2e-05 & 9e-09 & 2e-13 & \\
\hline
 SINC & 0.90 & 2e-01 & 5e-06 & 2e-13 & 5e-14 & 1e-14 & 1e-14 & 0.0023972816 & 6e-07 & 3e-15 & 6e-16 & 6e-16 & 6e-16 & 6e-16 & 0.0693682968 \\
 COS  &      & 5e-01 & 2e-02 & 1e-04 & 3e-07 & 2e-10 & 4e-14 & & 4e-01 & 2e-02 & 2e-04 & 6e-07 & 2e-10 & 7e-14 & \\
\hline
 SINC & 1.00 & 1e-02 & 7e-07 & 7e-14 & $\star$ & $\star$ & $\star$ & 0.0266495182 & 5e-08 & $\star$ & $\star$ & $\star$ & $\star$ & $\star$ & 0.4607202900 \\
 COS  &      & 1e-01 & 2e-03 & 2e-05 & 4e-08 & 2e-11 & 4e-15 & & 2e-02 & 1e-03 & 2e-05 & 6e-08 & 5e-11 & 1e-14 & \\
\hline
 SINC & 1.10 & 4e-03 & 2e-07 & 4e-15 & 1e-15 & $\star$ & $\star$ & 0.0949509784 & 3e-08 & 1e-16 & $\star$ & $\star$ & $\star$ & $\star$ & 0.9456809861 \\
 COS  &      & 6e-03 & 7e-04 & 2e-06 & 1e-08 & 6e-12 & $\star$ & & 5e-02 & 1e-04 & 2e-05 & 3e-08 & 3e-11 & 6e-15 & \\
\hline
 SINC & 1.20 & 2e-03 & 2e-08 & 2e-14 & 3e-15 & $\star$ & $\star$ & 0.1885055786 & 5e-08 & 2e-16 & $\star$ & $\star$ & $\star$ & $\star$ & 1.1723357082 \\
 COS  &      & 2e-02 & 4e-04 & 2e-06 & 2e-09 & 7e-13 & 1e-15 & & 1e-02 & 3e-04 & 1e-05 & 5e-08 & 4e-11 & 5e-15 & \\
\hline
 SINC & 1.30 & 4e-05 & 9e-08 & 3e-15 & 4e-16 & 1e-15 & 1e-15 & 0.2870818751 & 4e-08 & $\star$ & 2e-16 & 2e-16 & 2e-16 & 2e-16 & 1.2862729107 \\
 COS  &      & 4e-04 & 2e-04 & 2e-06 & 4e-09 & 3e-13 & 1e-15 & & 3e-02 & 8e-04 & 1e-06 & 4e-08 & 4e-11 & 4e-15 & \\
\hline
 SINC & 1.40 & 1e-03 & 6e-09 & 9e-15 & 6e-16 & $\star$ & $\star$ & 0.3860701385 & 3e-08 & 2e-16 & $\star$ & $\star$ & $\star$ & $\star$ & 1.3860485750 \\
 COS  &      & 6e-03 & 6e-06 & 1e-06 & 2e-09 & 1e-12 & $\star$ & & 6e-04 & 1e-03 & 8e-06 & 3e-08 & 3e-11 & 3e-16 & \\
\hline
\end{tabular}
\label{GBM_conv}
\end{table}

\noindent Then, two facts are immediately apparent from the table: \\
\textbf{Fact 1:} \emph{SINC} and COS experience different convergence rates, in general, and the type of considered option (either PV or a digital one) is going to change the patterns; \\
\textbf{Fact 2:} once the PDF truncation error is controlled and made sufficiently small, relatively low $\NCF$ is needed for the sum $\epsilon_2 + \epsilon_3$ to become negligible. This makes the \emph{SINC} candidate to be a very good approximation of the true option price. \\  

\noindent With particular respect to Fact 1, indeed, it is clear that the \emph{SINC} approach outperforms the COS when the stock price process follows a GBM. The convergence is much faster when dealing with PV put options on the left of Table \ref{GBM_conv} and the difference is even larger (far larger) when it comes to cash-or-nothing options on the right of the same table. \\

\noindent However, these figures are model-specific and cannot suffice to build up a complete idea about the convergence of the two methods. We therefore test them against other models in the following subsections and try to understand if we can extract something systematic about their convergence properties. In doing so, we take care that the models we select cover a large range of scenarios: from pure-diffusion models to infinite activity and rough volatility models.

\subsection{The Heston model}

Our second case of study is the celebrated Heston model. We report the dynamics of the price and volatility here to fix the notation:
\begin{align*}
& \dd S_t = S_t \sqrt{V_t} \dd B_t^S \\
& \dd V_t = -\lambda (V_t - \bar{V}) + \eta \sqrt{V_t} \dd B_t^V, \qquad \langle \dd B^S, \dd B^V \rangle_t = \rho.
\end{align*}
As with \cite{doi:10.1137/080718061}, again, we use parameters
\begin{align*}
S_0 = 1 \quad r & =0 \quad q=0 \quad T = \{0.1,1\} \quad \lambda=1.5768 \quad \eta = 0.5751 \\
& \bar{V}=0.0398 \quad V_0 = 0.0157 \quad \rho = -0.5711.
\end{align*}

\noindent Table \ref{Heston_conv_0.1} repeats the same analysis as before in the context of the Heston model (we only replace CoN with AoN options to show that the numbers are somehow invariant between the two). Expiration is in $0.1$ years, again, and we still observe improved convergence for the \emph{SINC} as opposed to the COS when dealing with put options. Once more, the difference is strikingly evident with digital options, where about four times as many evaluations of the CF are needed for the COS to replicate the same accuracy as the \emph{SINC}. \\

\begin{table}[h!]
\tiny \centering
\caption{Relative errors over PV and AoN put options for \emph{SINC} and COS at different values of $\NCF$ in the Heston model. Stars ($\star$) mean that the price fully conforms with the benchmark (up to the number of digits of the benchmark itself), straight lines denote relative errors that are larger than 100\%. [$T = 0.1$, $X_c = 2.0499$]}
\begin{tabular}{| l | c | c c c c c c |c| c c c c c c |c|}
 \hline
 & & \multicolumn{7}{c|}{PV put} & \multicolumn{7}{c|}{AoN put}\\
 \hline
 & & \multicolumn{6}{c|}{$\NCF$} & & \multicolumn{6}{c|}{$\NCF$} & \\
 \hline
 & K & 64 & 128 & 192 & 256 & 384 & 512 & benchmark & 64 & 128 & 192 & 256 & 384 & 512 & benchmark \\
 \hline
 SINC & 0.60 & \xdash & \xdash  & \xdash & 7e-01 & $\star$ & $\star$ & 0.0000000011 & \xdash & 4e-02 & $\star$ & $\star$ & $\star$ & $\star$ & 0.0000000397 \\
 COS  &      & \xdash & \xdash & \xdash & \xdash & 2e-01 & $\star$ & & \xdash &  \xdash & \xdash & \xdash & \xdash & 2e-02 & \\
\hline
 SINC & 0.70 & \xdash & \xdash  & 2e-02  & 2e-03 & $\star$  & $\star$ & 0.0000002363 & 4e-01 & 3e-04 & $\star$  & $\star$ & $\star$  & $\star$ & 0.0000077645\\
 COS  &      & \xdash & \xdash & \xdash & 1e-01 & 2e-03  & $\star$ & & \xdash & \xdash & \xdash & 2e-01 & 9e-04 & 2e-04 &  \\
\hline
 SINC & 0.80 & \xdash & 8e-02 & 3e-03 & 6e-05 & $\star$ & $\star$ & 0.0000198699 & 1e-02 & 6e-06 & $\star$ & $\star$ & $\star$ & $\star$ & 0.0006241401 \\
 COS  &      & \xdash & 3e-01 & 2e-02 & 1e-03 & 7e-06 & $\star$ & & \xdash &         7e-02 & 3e-02 & 1e-02 & 3e-04 & 4e-06 & \\
\hline
 SINC & 0.90 & 2e-01 & 3e-03 & 8e-05 & 2e-06 & $\star$ & $\star$ & 0.0008057899 & 8e-04 & 6e-07 & $\star$ & $\star$ & $\star$ & $\star$ & 0.0235660667 \\
 COS  &      & 2e-01 & 9e-03 & 7e-04 & 7e-05 & 9e-07 & $\star$ & & 3e-01 &        4e-02 & 5e-03 & 7e-04 & 2e-05 & 5e-07 & \\
\hline
 SINC & 1.00 & 2e-03 & 7e-05 & 4e-06 & 9e-08 & $\star$ & $\star$ & 0.0163700005 & 1e-04 & 1e-08 & 3e-10 & $\star$ & $\star$ & $\star$ & 0.4171058741 \\
 COS  &      & 2e-02 & 1e-03 & 1e-04 & 1e-06 & 2e-07 & 4e-09 & & 7e-02 &       5e-03 & 5e-04 & 1e-04 & 3e-06 & 8e-09 & \\
\hline
 SINC & 1.10 & 8e-04 & 5e-06 & 6e-07 & 2e-08 & $\star$ & $\star$ & 0.1000685530 & 9e-06 & 2e-08 & $\star$ & $\star$ & $\star$ & $\star$ & 0.9951985550 \\
 COS  &      & 2e-04 & 9e-05 & 1e-05 & 2e-06 & 3e-08 & $\star$ & & 2e-02 &       2e-03 & 2e-04 & 9e-06 & 5e-07 & 2e-08 & \\
\hline
 SINC & 1.20 & 8e-04 & 2e-05 & 4e-07 & 9e-09 & $\star$ & $\star$ & 0.2000001223 & 2e-05 & 1e-08 & $\star$ & $\star$ & $\star$ & $\star$ & 0.9999906356 \\
 COS  &      & 1e-03 & 1e-05 & 5e-06 & 2e-07 & 3e-09 & $\star$ & & 1e-03 &       1e-03 & 2e-05 & 2e-05 & 5e-07 & 1e-08 & \\
\hline
 SINC & 1.30 & 1e-04 & 1e-05 & 3e-08 & 7e-09 & $\star$ & $\star$ & 0.3000000002 & 1e-05 & 7e-09 & 1e-10 & $\star$ & $\star$ & $\star$ & 0.9999999797 \\
 COS  &      & 5e-04 & 3e-05 & 1e-06 & 8e-08 & 4e-09 & $\star$ & & 8e-03 & 4e-05 & 8e-05 & 2e-05 & 9e-08 & 9e-09 & \\
\hline
 SINC & 1.40 & 4e-04 & 4e-06 & 6e-08 & 5e-09 & $\star$ & $\star$ & 0.4000000000 & 3e-06 & 6e-09 & $\star$ & $\star$ & $\star$ & $\star$ & 0.9999999999 \\
 COS  &      & 1e-04 & 1e-05 & 1e-06 & 2e-07 & 2e-09 & $\star$ & & 8e-03 & 6e-04 & 6e-05 & 4e-06 & 1e-07 & 7e-09 & \\
\hline
\end{tabular}
\label{Heston_conv_0.1}
\end{table}

\noindent We also consider the case $T=1$ in Table \ref{Heston_conv_1} and observe that our considerations are basically unchanged regardless of the maturity of the option.

\begin{table}[h!]
\tiny \centering
\caption{Relative errors over PV and AoN put options for \emph{SINC} and COS at different values of $\NCF$ in the Heston model. Stars ($\star$) mean that the price fully conforms with the benchmark (up to the number of digits of the benchmark itself), straight lines denote relative errors that are larger than 100\%. [$T = 1$, $X_c = 12.1802$]}
\begin{tabular}{| l | c | c c c c c c |c| c c c c c c |c|}
 \hline
 & & \multicolumn{7}{c|}{PV put} & \multicolumn{7}{c|}{AoN put}\\
 \hline
 & & \multicolumn{6}{c|}{$\NCF$} & & \multicolumn{6}{c|}{$\NCF$} & \\
 \hline
 & K & 128 & 192 & 256 & 384 & 512 & 768 & benchmark & 128 & 192 & 256 & 384 & 512 & 768 & benchmark \\
 \hline
 SINC & 0.60 & 1e-01 & 1e-02 & 1e-03 & 4e-06 & 2e-07 & $\star$ & 0.0020880117 & 6e-04 & 3e-06 & 7e-08 & $\star$ & $\star$ & $\star$ & 0.0104586356 \\
 COS  &      & 8e-02 & 2e-02 & 1e-03 & 2e-05 & 1e-05 & 3e-07 & & 4e-01 & 4e-02 & 3e-02 & 3e-03 & 4e-04 & 3e-06 & \\
\hline
 SINC & 0.70 & 2e-02 & 4e-03 & 7e-04 & 3e-06 & 2e-07 & $\star$ & 0.0053328699 & 4e-04 & 4e-07 & 9e-08 & $\star$ & $\star$ & $\star$ & 0.0275972090 \\
 COS  &      & 1e-02 & 5e-03 & 2e-03 & 2e-04 & 8e-06 & 2e-07 & & 2e-01 & 6e-02 & 1e-02 & 6e-04 & 2e-04 & 2e-07 & \\
\hline
 SINC & 0.80 & 3e-02 & 8e-04 & 3e-04 & 2e-06 & 2e-08 & $\star$ & 0.0123663875 & 2e-04 & 2e-06 & 7e-09 & $\star$ & $\star$ & $\star$ & 0.0674328518 \\
 COS  &      & 3e-02 & 5e-04 & 1e-03 & 1e-04 & 1e-06 & 9e-08 & & 1e-01 &        4e-02 & 1e-02 & 2e-04 & 2e-04 & 2e-06 & \\
\hline
 SINC & 0.90 & 9e-03 & 9e-04 & 9e-05 & 1e-06 & 1e-08 & $\star$ & 0.0270953177 & 2e-05 & 7e-09 &       5e-09 & $\star$ & $\star$ & $\star$ & 0.1601233888 \\
 COS  &      &  3e-02 & 5e-04 & 1e-03 & 1e-04 & 9e-06 & 1e-07 & & 2e-02 & 3e-02 & 1e-03 & 1e-04 & 4e-06 & 2e-07 & \\
\hline
 SINC & 1.00 & 3e-03 & 6e-04 & 5e-07 & 5e-07 & 2e-08 & $\star$ & 0.0578515543 & 6e-05 & 1e-06 & 2e-08 & $\star$ & $\star$ & $\star$ & 0.3750835043 \\
 COS  &      & 9e-03 & 1e-03 & 8e-04 & 6e-05 & 3e-06 & $\star$ & & 6e-02 & 2e-02 & 3e-03 & 4e-04 & 6e-05 & 1e-06 & \\
\hline
 SINC & 1.10 & 2e-03 & 9e-05 & 2e-06 & 3e-07 & 8e-09 & $\star$ & 0.1178713500 & 5e-05 & 7e-07 & 6e-09 & $\star$ & $\star$ & $\star$ & 0.7236745029 \\
 COS  &      & 8e-03 & 1e-03 & 2e-04 & 5e-06 & 6e-07 & 3e-08 & & 4e-02 & 1e-02 &       4e-03 & 4e-04 & 5e-05 & 6e-07 & \\
\hline
 SINC & 1.20 & 1e-03 & 1e-04 & 1e-05 & 1e-07 & 2e-10 & $\star$ & 0.2048282813 & 3e-05 & 5e-07 & 7e-09 & $\star$ & $\star$ & $\star$ & 0.9222278418 \\
 COS  &      & 6e-03 & 1e-03 & 6e-05 & 2e-05 & 3e-07 & 9e-09 & & 8e-03 & 5e-03 & 3e-03 & 1e-05 & 3e-05 & 5e-07 & \\
\hline
 SINC & 1.30 & 1e-03 & 1e-04 & 2e-05 & 2e-07 & 4e-09 & $\star$ & 0.3014759365 & 2e-05 & 2e-07 & 2e-09 & $\star$ & $\star$ & $\star$ & 0.9771415014 \\
 COS  &      & 2e-03 & 8e-04 & 8e-05 & 7e-06 & 7e-07 & 1e-08 & &  2e-02 & 2e-03 & 2e-03 & 2e-04 & 2e-05 & 3e-07 & \\
\hline
 SINC & 1.40 & 9e-04 & 9e-05 & 2e-06 & 1e-07 & 3e-09 & $\star$ & 0.4005141485 & 5e-06 & 2e-07 & 4e-09 & $\star$ & $\star$ & $\star$ & 0.9922869299 \\
 COS  &      & 2e-03 & 5e-07 & 1e-04 & 1e-06 & 7e-07 & 3e-09 & & 9e-03 & 6e-03 & 1e-03 & 2e-04 & 4e-06 & 3e-07 & \\
\hline
\end{tabular}
\label{Heston_conv_1}
\end{table}

\subsection{CGMY}

As for infinite activity models we take CGMY as an example (recall that the Variance Gamma may be recovered as a special case). Given that parameter $Y$ is reported to play a special role in \cite{doi:10.1137/080718061}, we place ourselves in the exact same settings as they did, and select the following parameters for the model
\begin{align*}
S_0 = 1 \quad r = 0.1 \quad q = 0 \quad T = \{0.01,1\} \quad C = 1 \quad G = 5 \quad M = 5 \quad Y = \{0.50,1.50,1.98\}.
\end{align*}
We add the very short maturity $T=0.01$ to shed light on the case of short-dated out-of-the-money options. \\

\noindent Let us start with the case $Y=1.5$. We point the reader to Tables \ref{CGMY1.5_conv_1} and \ref{CGMY1.5_conv_0.01} so that he/she can realize that convergence of the methods follows similar patterns as we have commented for the GBM and the Heston model, with the usual distinction between PV and digital options. \\

\begin{table}[h!]
\tiny \centering
\caption{Relative errors over PV and CoN put options for \emph{SINC} and COS at different values of $\NCF$ with the CGMY process. Stars ($\star$) mean that the price fully conforms with the benchmark (up to the number of digits of the benchmark itself), straight lines denote relative errors that are larger than 100\%. [$Y=1.5$, $T=1$, $X_c=33.0891$]}
\begin{tabular}{| l | c | c c c c c c |c| c c c c c c |c|}
 \hline
 & & \multicolumn{7}{c|}{PV put} & \multicolumn{7}{c|}{CoN put}\\
 \hline
 & & \multicolumn{6}{c|}{$\NCF$} & & \multicolumn{6}{c|}{$\NCF$} & \\
 \hline
 & K & 16 & 32 & 48 & 64 & 96 & 128 & benchmark & 16 & 32 & 48 & 64 & 96 & 128 & benchmark \\
 \hline
 SINC & 0.60 & 4e-02 & 5e-05 & 1e-09 & $\star$ & $\star$ & $\star$ & 0.1703012777 & 2e-05 & $\star$ & $\star$ & $\star$ & $\star$ & $\star$ & 0.3030356173 \\
 COS  &      & 1e-01 & 1e-02 & 6e-04 & 1e-05 & 9e-10 & $\star$ & & 4e-02 & 6e-03 & 4e-04 & 2e-05 & 2e-09 & $\star$ & \\
\hline
 SINC & 0.70 & 5e-02 & 9e-08 & 6e-10 & $\star$ & $\star$ & $\star$ & 0.2230381363 & 2e-05 & $\star$ & $\star$ & $\star$ & $\star$ & $\star$ & 0.3838662218 \\
 COS  &      & 1e-01 & 8e-03 & 3e-04 & 4e-06 & 6e-11 & $\star$ & & 6e-02 & 9e-03 & 7e-04 & 2e-05 & 2e-09 & $\star$ & \\
\hline
 SINC & 0.80 & 5e-02 & 3e-05 & 2e-09 & $\star$ & $\star$ & $\star$ & 0.2797474350 & 2e-05 & $\star$ & $\star$ & $\star$ & $\star$ & $\star$ & 0.4678295710 \\
 COS  &      & 7e-02 & 4e-03 & 7e-05 & 2e-06 & 5e-10 & $\star$ & & 7e-02 & 1e-02 & 8e-04 & 2e-05 & 2e-09 & $\star$ & \\
\hline
 SINC & 0.90 & 5e-02 & 4e-05 & 2e-09 & $\star$ & $\star$ & $\star$ & 0.3398136494  & 2e-05 & $\star$ & $\star$ & $\star$ & $\star$ & $\star$ & 0.5541526890 \\
 COS  &      & 4e-02 & 1e-03 & 1e-04 & 7e-06 & 8e-10 & $\star$ & & 9e-02 & 1e-02 & 8e-04 & 2e-05 & 6e-10 & $\star$ & \\
\hline
 SINC & 1.00 & 4e-02 & 4e-05 & 2e-09 & $\star$ & $\star$ & $\star$ & 0.4027464727 & 2e-05 & $\star$ & $\star$ & $\star$ & $\star$ & $\star$ & 0.6422747911 \\
 COS  &      & 2e-02 & 1e-03 & 2e-04 & 1e-05 & 8e-10 & $\star$ & & 9e-02 & 1e-02 & 8e-04 & 2e-05 & 3e-10 & $\star$ & \\
\hline
 SINC & 1.10 & 4e-02 & 4e-05 & 2e-09 & $\star$ & $\star$ & $\star$ & 0.4681498500 & 1e-05 & $\star$ & $\star$ & $\star$ & $\star$ & $\star$ & 0.7317810268 \\
 COS  &      & 6e-03 & 3e-03 & 3e-04 & 1e-05 & 6e-10 & $\star$ & & 1e-01 & 1e-02 & 7e-04 & 1e-05 & 1e-09 & $\star$ & \\
\hline
 SINC & 1.20 & 4e-02 & 3e-05 & 1e-09 & $\star$ & $\star$ & $\star$ & 0.5356998308 & 6e-06 & $\star$ & $\star$ & $\star$ & $\star$ & $\star$ & 0.8223594103 \\
 COS  &      & 7e-03 & 4e-03 & 4e-04 & 1e-05 & 3e-10 & $\star$ & & 1e-01 & 1e-02 & 6e-04 & 6e-06 & 2e-09 & $\star$ & \\
\hline
 SINC & 1.30 & 3e-02 & 2e-05 & 1e-09 & $\star$ & $\star$ & $\star$ & 0.6051284646 & 8e-07 & $\star$ & $\star$ & $\star$ & $\star$ & $\star$ & 0.9137720089 \\
 COS  &      & 2e-02 & 5e-03 & 4e-04 & 1e-05 & 3e-11 & $\star$ & & 1e-01 & 1e-02 & 5e-04 & 1e-06 & 2e-09 & $\star$ & \\
\hline
 SINC & 1.40 & 3e-02 & 1e-05 & 1e-09 & $\star$ & $\star$ & $\star$ & 0.6762118959 & 4e-06 & $\star$ & $\star$ & $\star$ & $\star$ & $\star$ & 1.0058351838\\
 COS  &      & 3e-02 & 6e-03 & 4e-04 & 1e-05 & 2e-10 & $\star$ & & 1e-01 & 1e-02 & 4e-04 & 4e-06 & 2e-09 & $\star$ & \\
\hline
\end{tabular}
\label{CGMY1.5_conv_1}
\end{table}

\begin{table}[h!]
\tiny \centering
\caption{Relative errors over PV and CoN put options for \emph{SINC} and COS at different values of $\NCF$ with the CGMY process. Stars ($\star$) mean that the price fully conforms with the benchmark (up to the number of digits of the benchmark itself), straight lines denote relative errors that are larger than 100\%. [$Y=1.5$, $T=0.01$, $X_c=11.4582$]}
\begin{tabular}{| l | c | c c c c c c |c| c c c c c c |c|}
 \hline
 & & \multicolumn{7}{c|}{PV put} & \multicolumn{7}{c|}{CoN put}\\
 \hline
 & & \multicolumn{6}{c|}{$\NCF$} & & \multicolumn{6}{c|}{$\NCF$} & \\
 \hline
 & K & 16 & 32 & 64 & 128 & 256 & 512 & benchmark & 16 & 32 & 64 & 128 & 256 & 512 & benchmark \\
 \hline
 SINC & 0.60 & \xdash & \xdash & \xdash & 1e-01 & $\star$ & $\star$ & 0.0000620913 & \xdash & \xdash & 5e-03 & 1e-08 & $\star$ & $\star$ & 0.0006501357 \\
 COS  &      & \xdash & \xdash & \xdash & \xdash & 1e-02 & $\star$ & & \xdash & \xdash & \xdash & \xdash & 6e-03 & 1e-08 & \\
\hline
 SINC & 0.70 & \xdash & \xdash & \xdash & 5e-02 & 1e-07 & $\star$ & 0.0003572838 & \xdash & 8e-01 & 4e-03 & $\star$ & $\star$ & $\star$ & 0.0044404620 \\
 COS  &      & \xdash & \xdash & \xdash & 6e-01 & 3e-03 & $\star$ & & \xdash & \xdash & \xdash & 8e-01 & 4e-03 & $\star$ & \\
\hline
 SINC & 0.80 & \xdash & \xdash & 8e-01 & 1e-03 & $\star$ & $\star$ & 0.0021896005 & 8e-01 & 2e-01 & 1e-03 & $\star$ & $\star$ & $\star$ & 0.0320163483 \\
 COS  &      & \xdash & \xdash & \xdash & 6e-02 & 2e-04 & $\star$ & & \xdash & \xdash & 8e-01 & 2e-01 & 1e-03 & $\star$ & \\
\hline
 SINC & 0.90 & \xdash & 8e-01 & 3e-02 & 2e-03 & $\star$ & $\star$ & 0.0124149741 & 4e-01 & 5e-02 & 2e-04 & 7e-10 & $\star$ & $\star$ & 0.1764020073 \\
 COS  &      & \xdash & \xdash & 3e-01 & 2e-02 & 1e-04 & $\star$ & & \xdash & 9e-01 & 4e-01 & 5e-02 & 2e-04 & 7e-10 & \\
\hline
 SINC & 1.00 & 6e-01 & 3e-01 & 5e-02 & 5e-04 & 1e-09 & $\star$ & 0.0475882889 & 2e-02 & 3e-03 & 4e-05 & 3e-11 & $\star$ & $\star$ & 0.5239991369 \\
 COS  &      & \xdash & 8e-01 & 2e-01 & 1e-02 & 4e-05 & $\star$ & & 4e-02  & 3e-02 & 2e-02 & 3e-03 & 4e-05 & 3e-11 & \\
\hline
 SINC & 1.10 & 3e-01 & 1e-01 & 1e-03 & 2e-04 & $\star$ & $\star$ & 0.1156677674 &  1e-01 & 1e-02 & 6e-05 & 1e-10 & $\star$ & $\star$ & 0.8929163078 \\
 COS  &      & 8e-01 & 2e-01 & 2e-02 & 3e-03 & 1e-05 & $\star$ & & 3e-01 & 2e-01 & 1e-01 & 1e-02 & 6e-05 & 1e-10 & \\
\hline
 SINC & 1.20 & 2e-01 & 1e-02 & 1e-02 & 1e-04 & $\star$ & $\star$ & 0.2043804386 & 5e-02 & 7e-03 & 5e-05 & 1e-10 & $\star$ & $\star$ & 1.1281710458 \\
 COS  &      & 3e-01 & 2e-02 & 3e-02 & 2e-03 & 7e-06 & $\star$ & & 3e-01 & 2e-01 & 5e-02 & 7e-03 & 5e-05 & 1e-10 & \\
\hline
 SINC & 1.30 & 1e-01 & 3e-02 & 2e-03 & 8e-05 & $\star$ & $\star$ & 0.3008048296 & 6e-03 & 5e-03 & 4e-05 & 3e-11 & $\star$ & $\star$ & 1.2743502846 \\
 COS  &      & 9e-02 & 4e-02 & 3e-02 & 1e-03 & 6e-06 & $\star$ & & 3e-01 & 2e-01 & 6e-03 & 5e-03 & 4e-05 & 3e-11 & \\
\hline
 SINC & 1.40 & 6e-02 & 5e-02 & 7e-03 & 5e-05 & 2e-11 & $\star$ & 0.3995376021 & 4e-02 & 4e-03 & 2e-05 & 1e-10 & $\star$ & $\star$ & 1.3889715861 \\
 COS  &      & 2e-03 & 6e-02 & 1e-02 & 1e-03 & 5e-06 & $\star$ & & 3e-01 & 9e-02 & 4e-02 & 4e-03 & 2e-05 & 1e-10 & \\
\hline
\end{tabular}
\label{CGMY1.5_conv_0.01}
\end{table}

\noindent Then, approaching the limit case $Y=2$ does not introduce any issue (this was known with the COS method and \emph{SINC} is no different). Tables \ref{CGMY1.98_conv_1} and \ref{CGMY1.98_conv_0.01} confirm this fact. \\

\begin{table}[h!]
\tiny \centering
\caption{Relative errors over PV and CoN put options for \emph{SINC} and COS at different values of $\NCF$ with the CGMY process. Stars ($\star$) mean that the price fully conforms with the benchmark (up to the number of digits of the benchmark itself), straight lines denote relative errors that are larger than 100\%. [$Y=1.98$, $T=1$, $X_c=248.9047$]}
\begin{tabular}{| l | c | c c c c c c |c| c c c c c c |c|}
 \hline
 & & \multicolumn{7}{c|}{PV put} & \multicolumn{7}{c|}{CoN put}\\
 \hline
 & & \multicolumn{6}{c|}{$\NCF$} & & \multicolumn{6}{c|}{$\NCF$} & \\
 \hline
 & K & 16 & 32 & 48 & 64 & 96 & 128 & benchmark & 16 & 32 & 48 & 64 & 96 & 128 & benchmark \\
 \hline
 SINC & 0.60 & 1e-02 & 1e-05 & 6e-10 & $\star$ & $\star$ & $\star$ & 0.5429017201 & 5e-06 & $\star$ & $\star$ & $\star$ & $\star$ & $\star$ & 0.5429020811 \\
 COS  &      & 2e-02 & 3e-03 & 1e-04 & 2e-06 & 2e-10 & $\star$ & & 1e-02 & 4e-03 & 9e-05 & 5e-06 & 3e-10 & $\star$ & \\
\hline
 SINC & 0.70 & 1e-02 & 1e-05 & 4e-10 & $\star$ & $\star$ & $\star$ & 0.6333854028 & 5e-06 & $\star$ & $\star$ & $\star$ & $\star$ & $\star$ & 0.6333857942 \\
 COS  &      & 2e-02 & 3e-03 & 1e-04 & 3e-06 & $\star$ & $\star$ & & 1e-02 & 4e-03 & 8e-05 & 5e-06 & 2e-10 & $\star$ & \\
\hline
 SINC & 0.80 & 1e-02 & 1e-05 & 5e-10 & $\star$ & $\star$ & $\star$ & 0.7238690896 & 5e-06 & $\star$ & $\star$ & $\star$ & $\star$ & $\star$ & 0.7238695094 \\
 COS  &      & 2e-02 & 3e-03 & 1e-04 & 3e-06 & 1e-10 & $\star$ & & 1e-02 & 4e-03 & 7e-05 & 5e-06 & 2e-10 & $\star$ & \\
\hline
 SINC & 0.90 & 1e-02 & 1e-05 & 4e-10 & $\star$ & $\star$ & $\star$ & 0.8143527799 & 5e-06 & $\star$ & $\star$ & $\star$ & $\star$ & $\star$ & 0.8143532263 \\
 COS  &      & 1e-02 & 3e-03 & 1e-04 & 3e-06 & 1e-10 & $\star$ & & 1e-02 & 4e-03 & 7e-05 & 5e-06 & 2e-10 & $\star$ & \\
\hline
 SINC & 1.00 & 9e-03 & 1e-05 & 4e-10 & $\star$ & $\star$ & $\star$ & 0.9048364731 & 5e-06 & $\star$ & $\star$ & $\star$ & $\star$ & $\star$ & 0.9048369446 \\
 COS  &      & 1e-02 & 4e-03 & 1e-04 & 3e-06 & $\star$ & $\star$ & & 1e-02 & 4e-03 & 6e-05 & 5e-06 & 2e-10 & $\star$ & \\
\hline
 SINC & 1.10 & 9e-03 & 1e-05 & 4e-10 & $\star$ & $\star$ & $\star$ & 0.9953201687 & 5e-06 & $\star$ & $\star$ & $\star$ & $\star$ & $\star$ & 0.9953206642 \\
 COS  &      & 1e-02 & 4e-03 & 1e-04 & 3e-06 & 1e-10 & $\star$ & & 1e-02 & 4e-03 & 5e-05 & 5e-06 & 2e-10 & $\star$ & \\
\hline
 SINC & 1.20 & 8e-03 & 1e-05 & 3e-10 & $\star$ & $\star$ & $\star$ & 1.0858038665 & 5e-06 & $\star$ & $\star$ & $\star$ & $\star$ & $\star$ & 1.0858043849 \\
 COS  &      & 1e-02 & 4e-03 & 1e-04 & 3e-06 & 1e-10 & $\star$ & & 9e-03 & 4e-03 & 5e-05 & 5e-06 & 2e-10 & $\star$ & \\
\hline
 SINC & 1.30 & 8e-03 & 9e-06 & 4e-10 & $\star$ & $\star$ & $\star$ & 1.1762875661 & 5e-06 & $\star$ & $\star$ & $\star$ & $\star$ & $\star$ & 1.1762881065 \\
 COS  &      & 1e-02 & 4e-03 & 1e-04 & 3e-06 & 2e-10 & $\star$ & & 9e-03 & 4e-03 & 4e-05 & 5e-06 & 2e-10 & $\star$ & \\
\hline
 SINC & 1.40 & 8e-03 & 9e-06 & 3e-10 & $\star$ & $\star$ & $\star$ & 1.2667712674 & 5e-06 & $\star$ & $\star$ & $\star$ & $\star$ & $\star$ & 1.2667718289 \\
 COS  &      & 1e-02 & 4e-03 & 1e-04 & 3e-06 & 2e-10 & $\star$ & & 9e-03 & 4e-03 & 4e-05 & 5e-06 & 2e-10 & $\star$ & \\
\hline
\end{tabular}
\label{CGMY1.98_conv_1}
\end{table}

\begin{table}[h!]
\tiny \centering
\caption{Relative errors over PV and CoN put options for \emph{SINC} and COS at different values of $\NCF$ with the CGMY process. Stars ($\star$) mean that the price fully conforms with the benchmark (up to the number of digits of the benchmark itself), straight lines denote relative errors that are larger than 100\%. [$Y=1.98$, $T=0.01$, $X_c=24.9357$]}
\begin{tabular}{| l | c | c c c c c c |c| c c c c c c |c|}
 \hline
 & & \multicolumn{7}{c|}{PV put} & \multicolumn{7}{c|}{CoN put}\\
 \hline
 & & \multicolumn{6}{c|}{$\NCF$} & & \multicolumn{6}{c|}{$\NCF$} & \\
 \hline
 & K & 16 & 32 & 48 & 64 & 96 & 128 & benchmark & 16 & 32 & 48 & 64 & 96 & 128 & benchmark \\
 \hline
 SINC & 0.60 & 4e-02 & 5e-05 & $\star$ & $\star$ & $\star$ & $\star$ & 0.1359287091 & 2e-06 & $\star$ & $\star$ & $\star$ & $\star$ & $\star$ & 0.2916287066 \\
 COS  &      & 2e-01 & 2e-02 & 6e-04 & 1e-05 & $\star$ & $\star$ & & 9e-03 & 1e-03 & 9e-05 & 2e-06 & 3e-10 & $\star$ & \\
\hline
 SINC & 0.70 & 5e-02 & 2e-06 & 9e-10 & $\star$ & $\star$ & $\star$ & 0.1877507817 & 7e-06 & $\star$ & $\star$ & $\star$ & $\star$ & $\star$ & 0.3840967883 \\
 COS  &      & 2e-01 & 1e-02 & 4e-04 & 7e-06 & 6e-10 & $\star$ & & 3e-02 & 4e-03 & 3e-04 & 7e-06 & 7e-11 & $\star$ & \\
\hline
 SINC & 0.80 & 5e-02 & 3e-05 & $\star$ & $\star$ & $\star$ & $\star$ & 0.2453606494 & 1e-05 & $\star$ & $\star$ & $\star$ & $\star$ & $\star$ & 0.4816538267 \\
 COS  &      & 1e-01 & 8e-03 & 2e-04 & 2e-06 & $\star$ & $\star$ & & 5e-02 & 8e-03 & 5e-04 & 1e-05 & 2e-10 & $\star$ & \\
\hline
 SINC & 0.90 & 4e-02 & 4e-05 & $\star$ & $\star$ & $\star$ & $\star$ & 0.3079042304 & 1e-05 & $\star$ & $\star$ & $\star$ & $\star$ & $\star$ & 0.5828557330 \\
 COS  &      & 8e-02 & 4e-03 & 5e-05 & 1e-06 & $\star$ & $\star$ & & 7e-02 & 1e-02 & 6e-04 & 1e-05 & 2e-10 & $\star$ & \\
\hline
 SINC & 1.00 & 4e-02 & 3e-05 & $\star$ & $\star$ & $\star$ & $\star$ & 0.3746672106 & 1e-05 & $\star$ & $\star$ & $\star$ & $\star$ & $\star$ & 0.6866524319 \\
 COS  &      & 5e-02 & 1e-03 & 8e-05 & 3e-06 & $\star$ & $\star$ & & 8e-02 & 1e-02 & 5e-04 & 1e-05 & 3e-11 & $\star$ & \\
\hline
 SINC & 1.10 & 3e-02 &  3e-05 & 9e-11 & $\star$ & $\star$ & $\star$ & 0.4450537289 & 6e-06 & $\star$ & $\star$ & $\star$ & $\star$ & $\star$ & 0.7922765311\\
 COS  &      & 3e-02 & 8e-04 & 2e-04 & 5e-06 & 1e-10 & $\star$ & & 9e-02 & 1e-02 & 5e-04 & 7e-06 & 2e-10 & $\star$ & \\
\hline
 SINC & 1.20 & 3e-02 & 2e-05 & $\star$ & $\star$ & $\star$ & $\star$ & 0.5185661093 & 3e-06 & $\star$ & $\star$ & $\star$ & $\star$ & $\star$ & 0.8991643649 \\
 COS  &      & 1e-02 & 2e-03 & 2e-04 & 5e-06 & 5e-11 & $\star$ & & 1e-01 & 1e-02 & 4e-04 & 3e-06 & 3e-10 & $\star$ & \\
\hline
 SINC & 1.30 & 2e-02 & 9e-06 & 2e-10 & $\star$ & $\star$ & $\star$ & 0.5947873571 & 2e-07 & $\star$ & $\star$ & $\star$ & $\star$ & $\star$ & 1.0069001160 \\
 COS  &      & 1e-04 & 4e-03 & 3e-04 & 5e-06 & $\star$ & $\star$ & & 1e-01 & 1e-02 & 3e-04 & 1e-07 & 3e-10 & $\star$ & \\
\hline
 SINC & 1.40 & 2e-02 & 1e-06 & 2e-10 & $\star$ & $\star$ & $\star$ & 0.6733666332 & 3e-06 & $\star$ & $\star$ & $\star$ & $\star$ & $\star$ & 1.1151760866 \\
 COS  &      & 1e-02 & 5e-03 & 3e-04 & 4e-06 & $\star$ & $\star$ & & 1e-01 & 1e-02 & 2e-04 & 3e-06 & 3e-10 & $\star$ & \\
\hline
\end{tabular}
\label{CGMY1.98_conv_0.01}
\end{table}

\begin{figure}[h!]
	\centering
	\caption{PDF of the asset log-price under CGMY process for $T=1$ (lhs) and $T=0.01$ (rhs). Parameters: $C=1,G=5,M=5,Y=0.5$.}
	\begin{subfigure}[b]{0.49\textwidth}
		\centering
		\includegraphics[width=\textwidth]{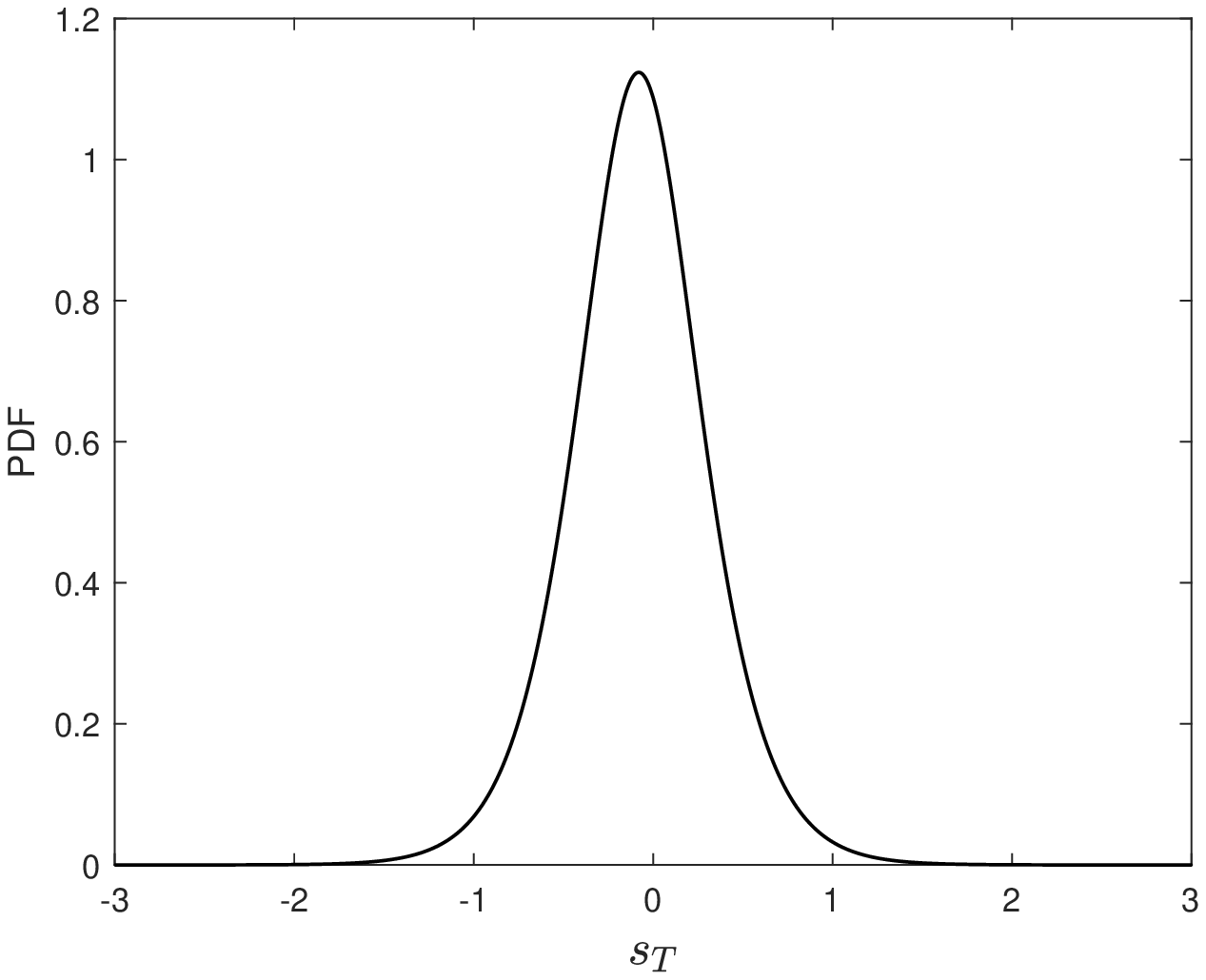}
	\end{subfigure}
	\begin{subfigure}[b]{0.49\textwidth}
		\centering
		\includegraphics[width=\textwidth]{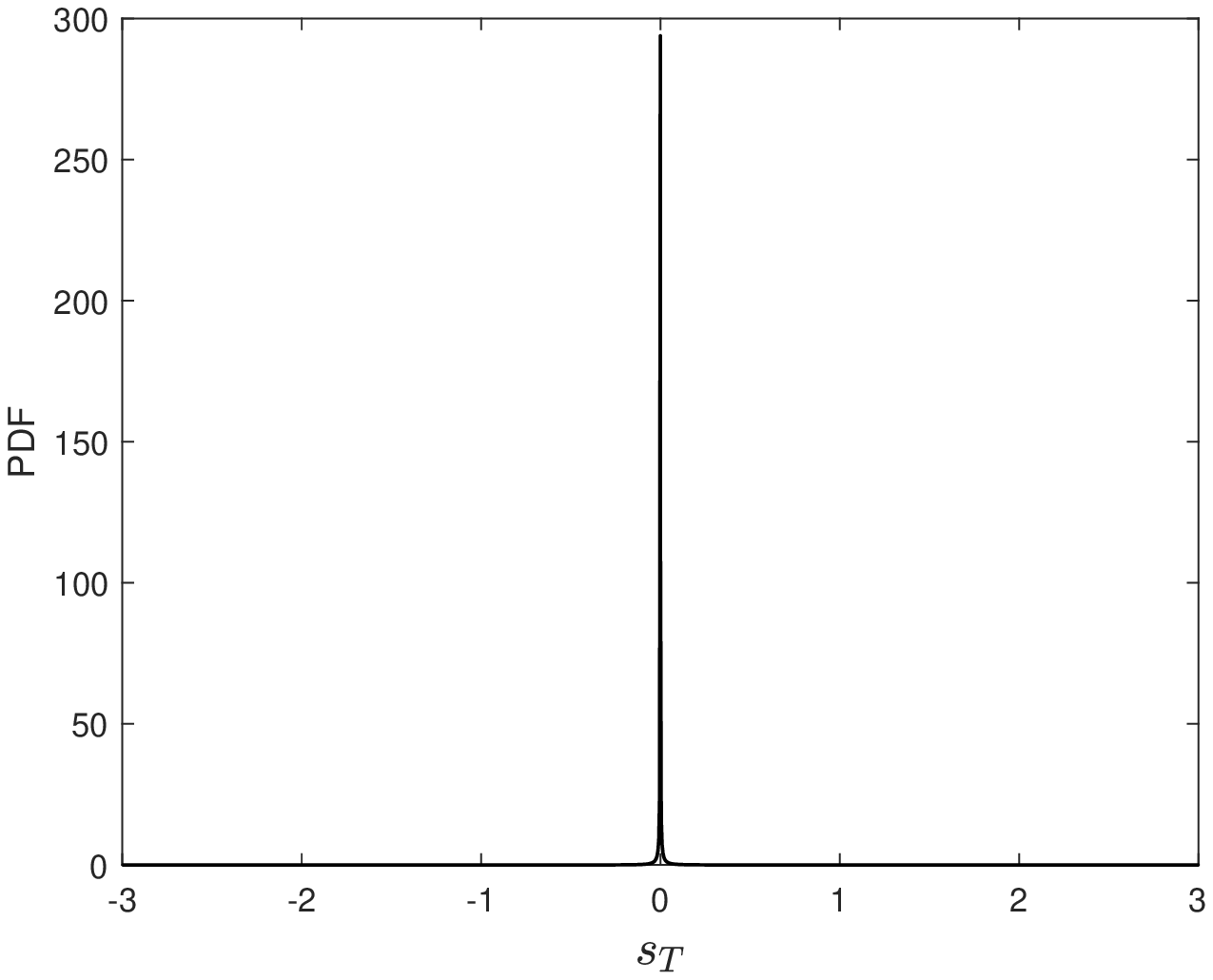}
	\end{subfigure}
	\label{pdf_CGMY}
\end{figure}

\noindent More involved is the case $Y=0.5$. Long maturities cause no deviations from the story we have told so far: look at Table \ref{CGMY0.5_conv_1} to see this. As expected, pricing short maturity options gets more complicated. When we look at PV put options on the lhs of Table \ref{CGMY0.5_conv_0.01} we see that the situation hints for an advantage of the COS. Only in this case we find the \emph{SINC} approach to exhibit poorer performance with respect to its competitor. Nevertheless, two things are worth noticing: (i) \emph{SINC} gets back its superiority (although less marked) when considering digital options (see rhs of Table \ref{CGMY0.5_conv_0.01}) and (ii) convergence is very slow for the COS as well as a result of enormously peaked density for the asset log-price. We plot the PDF for a CGMY process with $C=1,G=5,M=5,Y=0.5$ and maturity $T=1,T=0.01$ in Figure (\ref{pdf_CGMY}). Plotting the PDF is both very simple and computationally convenient when using the \emph{SINC} approach. The duality of the PDF and the CF is the theoretical backbone behind our procedure and the FT/FFT/frFFT provide the numerical tools for Fourier inversion. \\

\begin{table}[h!]
\tiny \centering
\caption{Relative errors over PV and CoN put options for \emph{SINC} and COS at different values of $\NCF$ with the CGMY process. Stars ($\star$) mean that the price fully conforms with the benchmark (up to the number of digits of the benchmark itself), straight lines denote relative errors that are larger than 100\%. [$Y=0.5$, $T=1$, $X_c=18.3512$]}
\begin{tabular}{| l | c | c c c c c c |c| c c c c c c |c|}
 \hline
 & & \multicolumn{7}{c|}{PV put} & \multicolumn{7}{c|}{CoN put}\\
 \hline
 & & \multicolumn{6}{c|}{$\NCF$} & & \multicolumn{6}{c|}{$\NCF$} & \\
 \hline
 & K & 16 & 32 & 64 & 128 & 256 & 512 & benchmark & 16 & 32 & 64 & 128 & 256 & 512 & benchmark \\
 \hline
 SINC & 0.60 & 5e-01 & 7e-01 & 2e-02 & 3e-05 & $\star$ & $\star$ & 0.0083082000 &  5e-02 & 8e-04 & 8e-07 & $\star$ & $\star$ & $\star$ & 0.0453996882 \\
 COS  &      & \xdash & 1e-01 & 1e-01 & 2e-03 & 2e-06 & $\star$ & & \xdash & 8e-01 & 5e-02 & 8e-04 & 8e-07 & $\star$ & \\
\hline
 SINC & 0.70 & \xdash & 2e-01 & 3e-03 & 9e-06 & $\star$ & $\star$ & 0.0189241836 & 4e-02 & 2e-03 & 2e-06 & $\star$ & $\star$ & $\star$ & 0.0984380619 \\
 COS  &      & \xdash & 3e-01 & 4e-02 & 2e-04 & 8e-07 & $\star$ & & \xdash & 5e-01 & 4e-02 & 2e-03 & 2e-06 & $\star$ & \\
\hline
 SINC & 0.80 & 8e-01 & 6e-03 & 4e-03 & 6e-06 & $\star$ & $\star$ & 0.0371703257 & 5e-02 & 1e-04 & 1e-06 & 2e-11 & $\star$ & $\star$ & 0.1819799859 \\
 COS  &      & \xdash & 4e-01 & 5e-03 & 6e-04 & 6e-07 & $\star$ & & 6e-01 & 3e-01 & 5e-02 & 1e-04 & 1e-06 & 2e-11 & \\
\hline
 SINC & 0.90 & 5e-01 & 6e-02 & 1e-04 & 5e-06 & $\star$ & $\star$ & 0.0648967583 & 3e-02 & 1e-03 & 3e-07 & 6e-11 & $\star$ & $\star$ & 0.2956669642 \\
 COS  &      & \xdash & 3e-01 & 2e-02 & 4e-05 & 5e-07 & $\star$ & & 2e-01 & 1e-01 & 3e-02 & 1e-03 & 3e-07 & 6e-11 & \\
\hline
 SINC & 1.00 & 4e-01 & 6e-02 & 2e-03 & 3e-06 & $\star$ & $\star$ & 0.1029669064 & 4e-03 & 2e-04 & 8e-07 & 3e-10 & $\star$ & $\star$ & 0.4323807669 \\
 COS  &      & 9e-01 & 2e-01 & 2e-02 & 3e-04 & 3e-07 & $\star$ & & 3e-02 & 2e-02 & 4e-03 & 2e-04 & 8e-07 & 3e-10 & \\
\hline
 SINC & 1.10 & 3e-01 & 5e-02 & 1e-03 & 4e-07 & $\star$ & $\star$ & 0.1511107358 & 1e-02 & 5e-04 & 1e-06 & $\star$ & $\star$ & $\star$ & 0.5813874463 \\
 COS  &      & 6e-01 & 1e-01 & 1e-02 & 1e-04 & 6e-08 & $\star$ & & 9e-02 & 5e-02 & 1e-02 & 5e-04 & 1e-06 & $\star$ & \\
\hline
 SINC & 1.20 & 2e-01 & 3e-02 & 9e-05 & 2e-06 & $\star$ & $\star$ & 0.2082023058 & 2e-02 & 5e-04 & 8e-07 & 3e-11 & $\star$ & $\star$ & 0.7328465863 \\
 COS  &      & 4e-01 & 9e-02 & 4e-03 & 8e-05 & 1e-07 & $\star$ & & 2e-01 & 8e-02 & 2e-02 & 5e-04 & 8e-07 & 3e-11 & \\
\hline
 SINC & 1.30 & 2e-01 & 2e-02 & 5e-04 & 5e-07 & $\star$ & $\star$ & 0.2727258052 & 2e-02 & 5e-05 & 5e-07 & $\star$ & $\star$ & $\star$ & 0.8803504028 \\
 COS  &      & 3e-01 & 4e-02 & 2e-03 & 1e-04 & 1e-07 & $\star$ & & 2e-01 & 1e-01 & 2e-02 & 5e-05 & 5e-07 & $\star$ & \\
\hline
 SINC & 1.40 & 1e-01 & 4e-03 & 6e-04 & 1e-06 & 4e-11 & $\star$ & 0.3431572893 & 1e-02 & 3e-04 & 8e-07 & 3e-11 & $\star$ & $\star$ & 1.0208938948 \\
 COS  &      & 2e-01 & 1e-02 & 4e-03 & 7e-05 & 4e-08 & $\star$ & & 2e-01 & 1e-01 & 1e-02 & 3e-04 & 8e-07 & 3e-11 & \\
\hline
\end{tabular}
\label{CGMY0.5_conv_1}
\end{table}

\begin{table}[h!]
\tiny \centering
\caption{Relative errors over PV and CoN put options for \emph{SINC} and COS at different values of $\NCF$ with the CGMY process. Stars ($\star$) mean that the price fully conforms with the benchmark (up to the number of digits of the benchmark itself), straight lines denote relative errors that are larger than 100\%. [$Y=0.5$, $T=0.01$, $X_c=12.0723$]}
\begin{tabular}{| l | c | c c c c c c |c| c c c c c c |c|}
 \hline
 & & \multicolumn{7}{c|}{PV put} & \multicolumn{7}{c|}{CoN put}\\
 \hline
 & & \multicolumn{6}{c|}{$\NCF$} & & \multicolumn{6}{c|}{$\NCF$} & \\
 \hline
 & K & 256 & 512 & 1024 & 2048 & 4096 & 8192 & benchmark & 256 & 512 & 1024 & 2048 & 4096 & 8192 & benchmark \\
 \hline
 SINC & 0.60 & \xdash & \xdash & \xdash & \xdash & 8e-01 & 5e-01 & 0.0000223408 & \xdash & \xdash & 8e-01 & 2e-01 & 8e-02 & 3e-03 & 0.0001785593 \\
 COS  &      & \xdash & \xdash & 6e-01 & 1e-01 & 2e-02 & 3e-03 & & \xdash & \xdash & \xdash & \xdash & 8e-01 & 2e-01 & \\
\hline
 SINC & 0.70 & \xdash & \xdash & \xdash & 3e-01 & 7e-01 & 2e-01 & 0.0000799072 & \xdash & \xdash & 8e-02 & 2e-01 & 4e-02 & 5e-03 & 0.0006873067 \\
 COS  &      & \xdash & \xdash & 9e-02 & 4e-02 & 1e-02 & 3e-04 & & \xdash & \xdash & \xdash & \xdash & 8e-02 & 2e-01 & \\
\hline
 SINC & 0.80 & \xdash & 6e-01 & \xdash & 7e-01 & 2e-01 & 1e-02 & 0.0002674771 & 3e-01 & 8e-01 & 3e-01 & 7e-02 & 4e-03 & 4e-03 & 0.0025728961 \\
 COS  &      & \xdash & 6e-01 & 2e-01 & 8e-03 & 2e-03 & 6e-04 & & \xdash & \xdash & 3e-01 & 8e-01 & 3e-01 & 7e-02 & \\
\hline
 SINC & 0.90 & \xdash & 2e-01 & 6e-01 & 2e-01 & 6e-02 & 7e-03 & 0.0009262546 & 2e-01 & 5e-01 & 2e-01 & 4e-02 & 5e-03 & 2e-03 & 0.0110260134 \\
 COS  &      & 2e-01 & 2e-01 & 1e-01 & 5e-03 & 1e-03 & 3e-04 & & \xdash & \xdash & 2e-01 & 5e-01 & 2e-01 & 4e-02 & \\
\hline
 SINC & 1.00 & 1e-01 & 8e-02 & 4e-02 & 2e-02 & 6e-03 & 2e-03 & 0.0060510208 & 1e-01 & 9e-02 & 8e-02 & 6e-02 & 3e-02 & 1e-02 & 0.4459496554 \\
 COS  &      & \xdash & 5e-01 & 2e-01 & 6e-02 & 2e-02 & 5e-03 & & 1e-01 & 1e-01 & 1e-01 & 9e-02 & 8e-02 & 6e-02 & \\
\hline
 SINC & 1.10 & 4e-02 & 2e-02 & 6e-03 & 2e-03 & 5e-04 & 2e-05 & 0.1004887993 & 2e-02 & 6e-03 & 2e-03 & 5e-04 & 2e-05 & 3e-05 & 1.0832677912 \\
 COS  &      & 1e-02 & 1e-03 & 9e-05 & 3e-05 & 2e-05 & 5e-06 & & 8e-02 & 4e-02 & 2e-02 & 6e-03 & 2e-03 & 5e-04 & \\
\hline
 SINC & 1.20 & 2e-02 & 5e-03 & 6e-04 & 1e-03 & 3e-04 & 2e-05 & 0.1995452121 & 5e-03 & 6e-04 & 1e-03 & 3e-04 & 2e-05 & 2e-05 & 1.1929282653 \\
 COS  &      & 4e-03 & 1e-03 & 3e-04 & 7e-05 & 5e-06 & 1e-06 & & 4e-02 & 2e-02 & 5e-03 & 6e-04 & 1e-03 & 3e-04 & \\
\hline
 SINC & 1.30 & 2e-03 & 6e-03 & 2e-03 & 7e-05 & 3e-04 & 6e-05 & 0.2991115246 & 6e-03 & 2e-03 & 7e-05 & 3e-04 & 6e-05 & 1e-05 & 1.2958046554 \\
 COS  &      & 3e-03 & 9e-04 & 5e-05 & 2e-05 & 7e-06 & 2e-07 & & 2e-02 & 2e-03 & 6e-03 & 2e-03 & 7e-05 & 3e-04 & \\
\hline
 SINC & 1.40 & 1e-02 & 3e-03 & 1e-04 & 6e-04 & 2e-04 & 4e-05 & 0.3988490890 & 3e-03 & 1e-04 & 6e-04 & 2e-04 & 4e-05 & 4e-06 & 1.3969698655 \\
 COS  &      & 3e-03 & 2e-04 & 8e-05 & 2e-05 & 4e-07 & 2e-07 & & 3e-03 & 1e-02 & 3e-03 & 1e-04 & 6e-04 & 2e-04 & \\
\hline
\end{tabular}
\label{CGMY0.5_conv_0.01}
\end{table}

\noindent We have the following obvious fact: \\
\textbf{Fact 3:} the number of evaluations of the CF that guarantees convergence to the benchmark increases with peaked density functions.

\subsection{The rough Heston model}\label{rHeston_SINCvsCOS}

A more fashionable example is finally given by the rHeston model. We recall it in the following for readers' convenience. \\

\noindent The (generalized) rough Heston model from~\cite{eleuch2018} is described by the following equations:
\begin{subequations} 
  \begin{align*}
    & \dd S_t = S_t \sqrt{V_t}\{\rho \dd B_t + \sqrt{1-\rho^2} \dd B_t^\perp\}, \\ 
    & V_t  = V_0 + \frac{\lambda}{\Gamma(H+\frac{1}{2})} \int_0^t \frac{\theta^0(s) - V_s}{(t-s)^{\frac{1}{2}-H}} \dd s + \frac{\nu}{\Gamma(H+\frac{1}{2})} \int_0^t \frac{\sqrt{V_s}}{(t-s)^{\frac{1}{2}-H}} \dd B_s, 
  \end{align*}
\end{subequations}
where $V_0$, $\lambda$, and $\nu$ are positive real numbers, $\rho \in [-1,1]$. The deterministic function $\theta^0(t)$ is positive and satisfies few constraints specified in~\cite{eleuch2018}. The coefficient $H \in (0,1/2]$ is shown to govern the smoothness of the volatility, whose trajectories enjoy H\"older continuity $H-\epsilon$ for any $\epsilon>0$. It is therefore clear that the choice $H<1/2$ allows for a rough behavior of the volatility process and the case $H=1/2$ amounts to the classical Heston model with time-dependent mean reversion level. \\

\noindent \cite{eleuch2018} proved also that the product $\lambda\theta^0(\cdot)$ is directly inferred from the time-$0$ forward variance curve $\xi_0(t) = \mathbb{E}[V_t|\mathcal{F}_0 ]= \mathbb{E}[V_t]$, leading to the following specification of the model for $\lambda\rightarrow 0$:
\begin{subequations} 
  \begin{align*}
    & \dd S_t = S_t \sqrt{V_t}\{\rho \dd B_t + \sqrt{1-\rho^2} \dd B_t^\perp\}, \\ 
    & V_t = \xi_0(t) + \frac{\nu}{\Gamma(H+\frac{1}{2})} \int_0^t \frac{\sqrt{V_s}}{(t-s)^{\frac{1}{2}-H}} \dd B_s. 
  \end{align*}
\end{subequations}

\noindent This is extremely convenient for calibration purposes thanks to the reduced dimensionality of the problem. We will consequently work under this last specification throughout the rest of the paper, thus placing ourselves in the same setting of \cite{el2019roughening}. Indeed our final point with the rHeston model will be to show that the SINC approach is a very efficient solution for calibration. This will be the content of the following section. \\ 

\noindent The forward variance curve is a state variable in the model and it also enters the CF of the asset log-price (see \cite{eleuch2018} for further details):
\begin{align*}
\varphi(a,t) = \mathbb{E}\bigg[ \exp\bigg\{ia\log\bigg(\frac{S_t}{S_0}\bigg)\bigg\} \bigg] = \exp \bigg( \int_0^t D^\alpha h(a,t-s)\xi_0(s) \dd s \bigg), 
\end{align*}
where $\alpha = H + \frac{1}{2}$, $h(a,t)$ is the unique continuous solution of the fractional Riccati equation
\begin{align}
D^\alpha h(a,t) = -\frac{1}{2}a(a+i) + ia\rho\nu h(a,t) + \frac{\nu^2}{2}h^2(a,t), \qquad I^{1-\alpha}h(a,0) = 0, \label{fracRic_rHeston}
\end{align}
and $D^\alpha$, $I^{1-\alpha}$ denote the Riemann-Liouville fractional derivative and fractional integral of order $\alpha$ and $1-\alpha$, respectively~\footnote{The Riemann-Liouville fractional derivative of a function $f$ is defined as 
\begin{align*}
D^{\alpha} f(t) = \frac{1}{\Gamma(1-\alpha)} \frac{\dd}{\dd t} \int_0^t (t-s)^{-\alpha} f(s) \dd s \qquad \alpha \in [0,1),
\end{align*}
provided that it exists. Similarly the fractional integral, provided that it exists, is given by
\begin{align*}
I^{\alpha} f(t) = \frac{1}{\Gamma(\alpha)} \int_0^t (t-s)^{\alpha-1}f(s) \dd s \qquad \alpha \in (0,1].
\end{align*}
}. \\

\noindent Now, Equation~(\ref{fracRic_rHeston}) is a rough version of the Riccati ODE which emerges in the classical Heston model with zero mean reversion. Here, the standard derivative is replaced by a fractional one. However, such a small change has relevant implications. The rHeston Riccati equation has no explicit solution and needs to be solved using numerical methods which are not really plain. We are not discussing the general issue of an efficient computation of the CF, which topic has been largely debated in the last few years (from a standard application of the Adams scheme of \cite{diethelm2004detailed} to more problem-specific techniques like the rational approximation of \cite{gatheral2019rational} and the hybrid method of \cite{callegaro2020fast}). We simply claim that, given any approximation to the CF, the \emph{SINC} is a very effective method to perform pricing and calibration. We will therefore stick with the rational approximation to the CF of \cite{gatheral2019rational} to compute the CF, and discuss our results within that setting. The interested reader is referred to the original paper for a complete discussion about the approximation; in particular, Equations (4.1) and (4.12)-(4.17) in \cite{gatheral2019rational} will do most of the job. \\

\noindent We use the following parameters for the experiments that follow
\begin{align*}
S_0 = 1 \quad r=0 \quad q=0 \quad T=\{0.01,1\} \quad H = 0.05 \quad \nu = 0.4 \quad \rho = -0.65,
\end{align*}  
and assume the forward variance curve is flat at $\xi_0(\cdot)=0.0256$. \\

\noindent Let us start with the typical example where expiration is in 1 year, i.e. $T=1$. As we always did throughout this section, we report convergence results for PV and digital put options (AoN in this case) in Table \ref{rHeston_conv_1}. We have often observed a gap in the performances of the \emph{SINC} and the COS method when dealing with PV options and we have learnt that the gap gets larger when moving to their digital components. This time it is not different. Therefore, it would be the case that we try to understand where such a behavior comes from and we refer to the following Figure (\ref{conv_long_fig}) in doing so \footnote{We only focus on options which are struck at $K=0.80$ but stress that the pattern does not depend on the moneyness (as one can guess from Table \ref{rHeston_conv_1}).}. \\

\begin{table}[h!]
\tiny \centering
\caption{Relative errors over PV and AoN put options for \emph{SINC} and COS at different values of $\NCF$ in the rough Heston model. Stars ($\star$) mean that the price fully conforms with the benchmark (up to the number of digits of the benchmark itself), straight lines denote relative errors that are larger than 100\%. [$T=1$, $X_c=18.9469$]}
\begin{tabular}{| l | c | c c c c c c |c| c c c c c c |c|}
 \hline
 & & \multicolumn{7}{c|}{PV put} & \multicolumn{7}{c|}{dig put}\\
 \hline
 & & \multicolumn{6}{c|}{$\NCF$} & & \multicolumn{6}{c|}{$\NCF$} & \\
 \hline
 & K & 256 & 512 & 768 & 1024 & 1536 & 2048 & benchmark & 256 & 512 & 768 & 1024 & 1536 & 2048 & benchmark \\
 \hline
 SINC & 0.60 & 2e-02 & 6e-03 & 2e-05 & 2e-06 & $\star$ & $\star$ & 0.003190745 & 4e-03 & 2e-06 & $\star$ & $\star$ & $\star$ & $\star$ & 0.0105749538 \\
 COS  &      & 8e-02 & 6e-03 & 3e-04 & 3e-05 & 1e-06 & $\star$ & & 2e-01 & 2e-02 &        1e-02 & 2e-03 & 2e-05 & 9e-07 & \\
\hline
 SINC & 0.70 & 4e-02 & 3e-03 & 6e-05 & 3e-07 & $\star$ & $\star$ & 0.006322036  & 2e-03 & 2e-07 & $\star$ & $\star$ & $\star$ & $\star$ & 0.0226378550 \\
 COS  &      & 6e-02 & 4e-03 & 1e-05 & 6e-05 & 3e-07 & $\star$ & & 9e-02 & 3e-02 & 1e-02 & 1e-03 & 4e-05 & 3e-08 & \\
\hline
 SINC & 0.80 & 1e-02 & 1e-03 & 3e-05 & 2e-07 & $\star$ & $\star$ & 0.011948775 & 2e-03 & 5e-07 & $\star$ & $\star$ & $\star$ & $\star$ & 0.0477997904 \\
 COS  &      & 5e-02 & 4e-03 & 4e-04 & 1e-05 & 9e-07 & $\star$ & & 1e-01 & 3e-03 & 5e-03 & 1e-03 & 2e-05 & 4e-07 & \\
\hline
 SINC & 0.90 & 3e-02 & 1e-03 & 2e-05 & 4e-07 & $\star$ & $\star$ & 0.022432028 & 1e-03 & 5e-07 & $\star$ & $\star$ & $\star$ & $\star$ & 0.1081020775 \\
 COS  &      & 4e-02 & 2e-03 & 3e-04 & 3e-05 & 6e-07 & $\star$ & & 2e-01 & 3e-02 & 6e-03 & 1e-03 & 2e-05 & 4e-07 & \\
\hline
 SINC & 1.00 & 2e-03 & 4e-04 & 8e-06 & 1e-07 & $\star$ & $\star$ & 0.045518977 & 7e-04 & 3e-07 & $\star$ & $\star$ & $\star$ & $\star$ & 0.3222614106 \\
 COS  &      & 2e-02 & 2e-03 & 4e-04 & 3e-05 & 9e-07 & $\star$ & & 1e-01 & 2e-02 & 2e-04 & 6e-04 & 7e-06 & 3e-07 & \\
\hline
 SINC & 1.10 & 3e-03 & 7e-05 & 5e-06 & 9e-08 & $\star$ & $\star$ & 0.108597244 & 3e-04 & 1e-08 & $\star$ & $\star$ & $\star$ & $\star$ & 0.8378414609 \\
 COS  &      & 1e-03 & 1e-03 & 1e-04 & 6e-06 & 3e-07 & 2e-10 & & 6e-02 & 3e-03  & 1e-03 & 3e-04 & 6e-06 & 4e-09 & \\
\hline
 SINC & 1.20 & 3e-03 & 8e-05 & 1e-06 & 3e-10 & $\star$ & $\star$ & 0.202190574 & 3e-05 & 4e-08 & $\star$ & $\star$ & $\star$ & $\star$ & 0.9673515242 \\
 COS  &      & 5e-03 & 5e-04 & 7e-05 & 9e-06 & 2e-07 & $\star$ & & 3e-02 & 4e-03 & 5e-04 & 5e-05 & 4e-07 & 4e-08 & \\
\hline
 SINC & 1.30 & 3e-03 & 8e-05 & 2e-06 & 3e-08 & $\star$ & $\star$ & 0.300785493 & 8e-05 & 3e-08 & $\star$ & $\star$ & $\star$ & $\star$ & 0.9898679762 \\
 COS  &      & 4e-03 & 2e-04 & 3e-05 & 3e-06 & 5e-08 & $\star$ & & 1e-02 &4e-03 & 4e-04 & 1e-04 & 2e-06 & 4e-08 & \\
\hline
 SINC & 1.40 & 2e-03 & 7e-05 & 2e-06 & 3e-08 & $\star$ & $\star$ & 0.400341035 & 7e-05 & 2e-08 & $\star$ & $\star$ & $\star$ & $\star$ & 0.9960163357 \\
 COS  &      & 1e-03 & 8e-05 & 1e-05 & 2e-06 & 3e-08 & $\star$ & & 2e-02 & 3e-03 & 6e-04 & 9e-05 & 2e-06 & 3e-08 & \\
\hline
\end{tabular}
\label{rHeston_conv_1}
\end{table}

\begin{figure}[h!]
	\centering
	\caption{Convergence of the \emph{SINC} (lhs) and the COS (rhs) method. Red (dashed), blue (dot-dashed), and black (bold) lines are the CoN, AoN, and put options, respectively. Light blue horizontal lines denote the benchmarks. $T=1$ and $K=0.80$.}
	\begin{subfigure}[b]{0.49\textwidth}
		\centering
		\includegraphics[width=\textwidth]{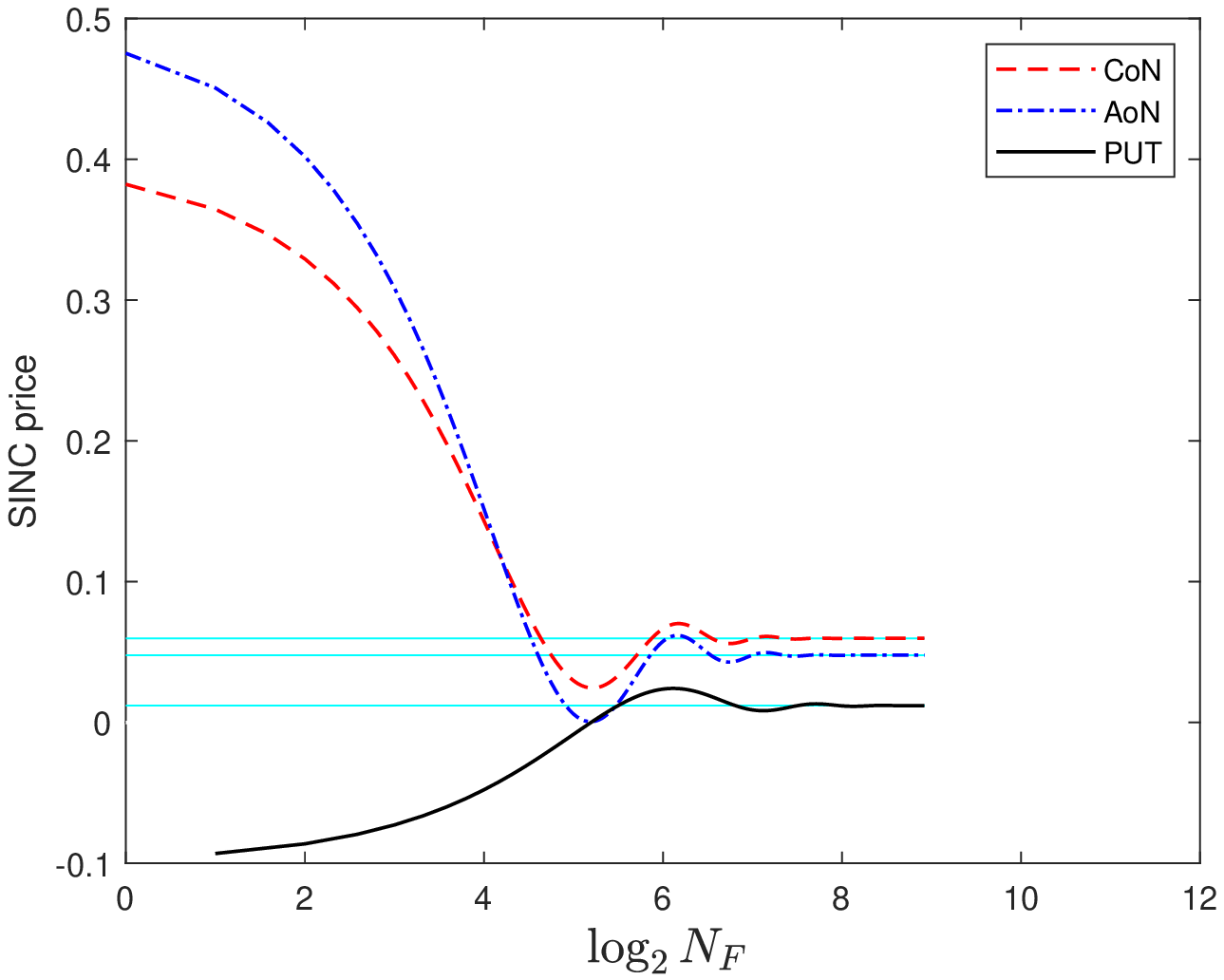}
	\end{subfigure}
	\begin{subfigure}[b]{0.49\textwidth}
		\centering
		\includegraphics[width=\textwidth]{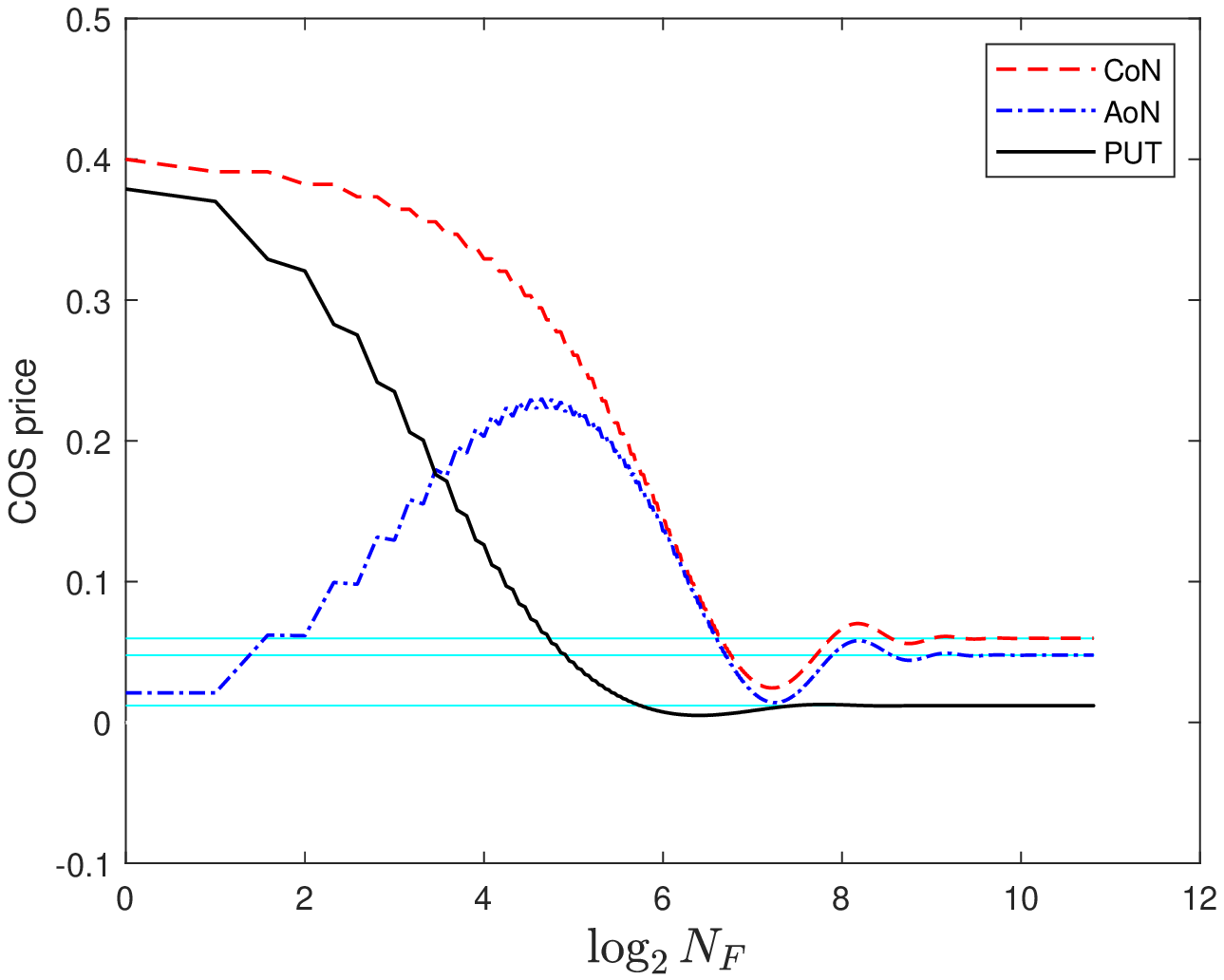}
	\end{subfigure}
	\label{conv_long_fig}
\end{figure}

\noindent The idea behind these charts is that we take our benchmarks as a reference, compute prices with the two methods by increasing $\NCF$ of one unit per time and stop when we have reached satisfactory accuracy (say 8 digits) on both CoN and AoN puts. Not surprisingly, then, the \emph{SINC} is shown to meet the target for much lower $\NCF$, and the oscillations on the digital options last longer on the rhs than the lhs of the figure. Also, digital options are less stable than PV puts, for both the methods, but the composition of CoN and AoN options introduces cancellations that are beneficial to the COS much more than they are to the \emph{SINC}. This, in fact, explains PV put prices within the COS to finally catch up the \emph{SINC}. \\

\begin{figure}[h!]
	\centering
	\caption{Convergence of the \emph{SINC} (lhs) and the COS (rhs) method. Red (dashed), blue (dot-dashed), and black (bold) lines are the CoN, AoN, and put options, respectively. Light blue horizontal lines denote the benchmarks. $T=0.01$ and $K=0.80$.}
	\begin{subfigure}[b]{0.49\textwidth}
		\centering
		\includegraphics[width=\textwidth]{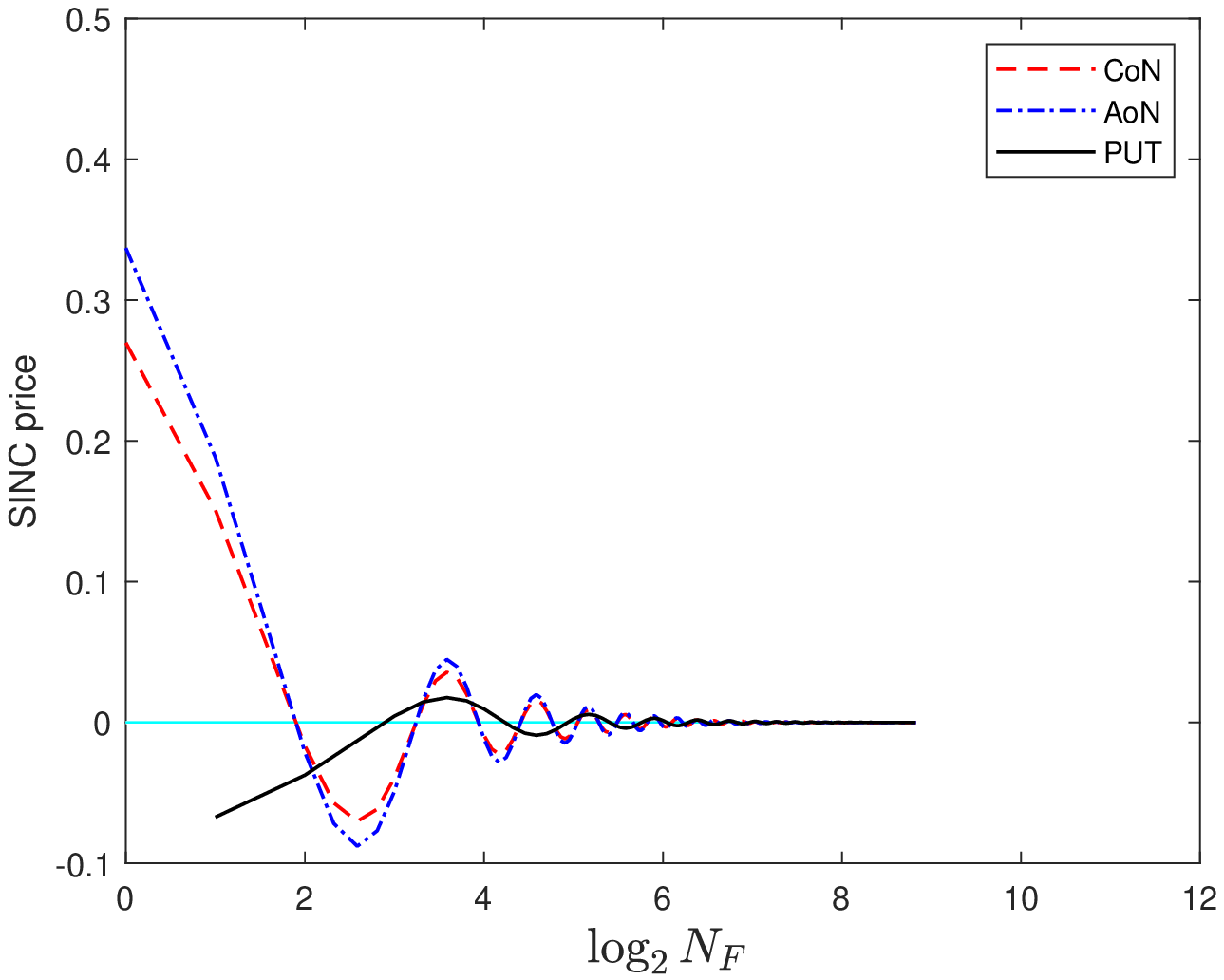}
	\end{subfigure}
	\begin{subfigure}[b]{0.49\textwidth}
		\centering
		\includegraphics[width=\textwidth]{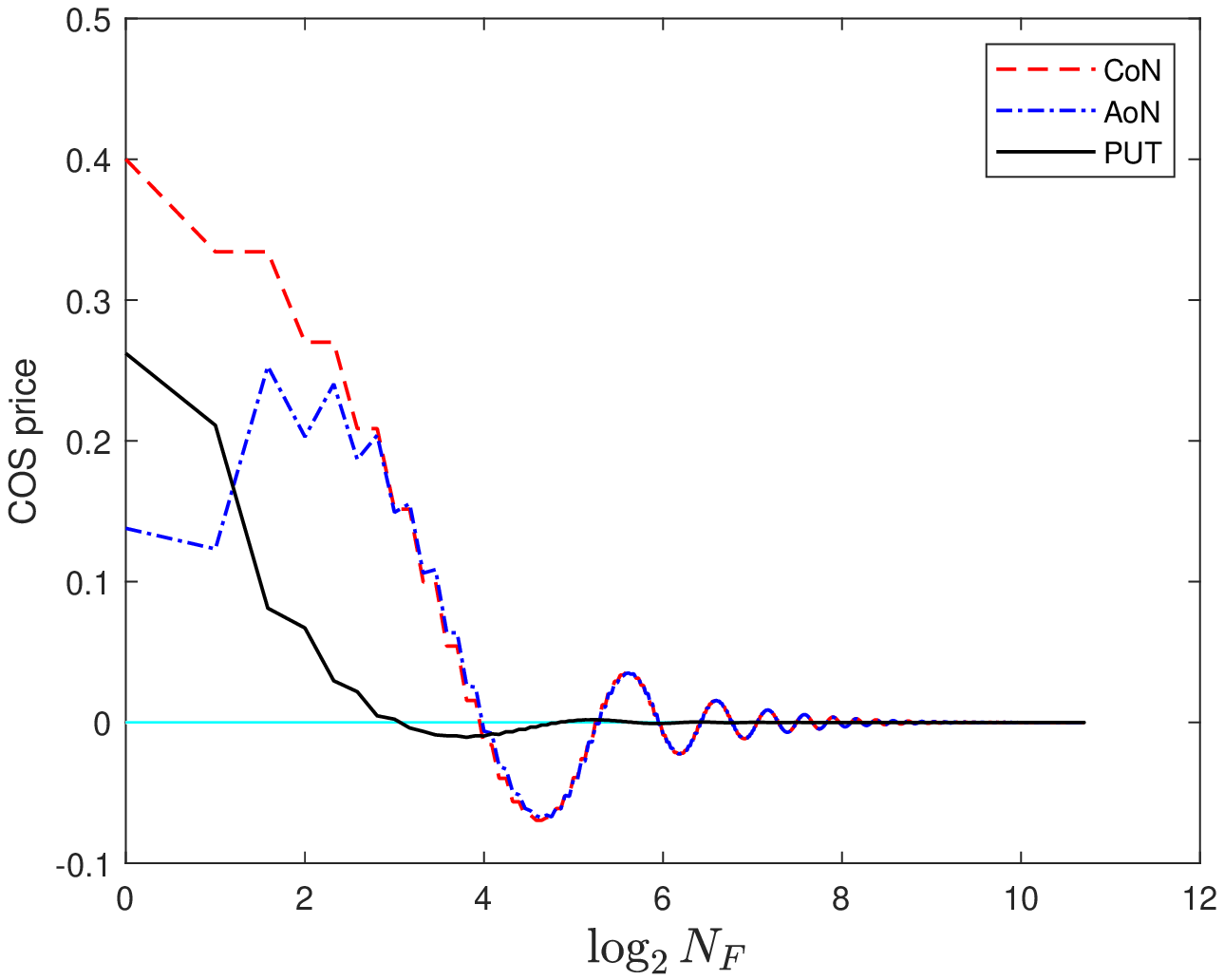}
	\end{subfigure}
	\label{conv_short_fig}
\end{figure}

\noindent Moreover, we also consider the challenging case where $T=0.01$. If this seems too short, it is still something one may encounter during the calibration process, and it goes in the direction of pricing the weekly options. We consequently produce Table \ref{rHeston_conv_0.01} and accompany it by Figure (\ref{conv_short_fig}) (where we study the case $K=0.80$ in more details). \\

\begin{table}[h!]
\tiny \centering
\caption{Relative errors over PV and AoN put options for \emph{SINC} and COS at different values of $\NCF$ in the rough Heston model. Stars ($\star$) mean that the price fully conforms with the benchmark (up to the number of digits of the benchmark itself), straight lines denote relative errors that are larger than 100\%. [$T= 0.01$, $X_c=2.7074$]}
\begin{tabular}{| l | c | c c c c c c |c| c c c c c c |c|}
 \hline
 & & \multicolumn{7}{c|}{PV put} & \multicolumn{7}{c|}{AoN put}\\
 \hline
 & & \multicolumn{6}{c|}{$\NCF$} & & \multicolumn{6}{c|}{$\NCF$} & \\
 \hline
 & K & 256 & 512 & 768 & 1024 & 1536 & 2048 & benchmark & 256 & 512 & 768 & 1024 & 1536 & 2048 & benchmark \\
 \hline
 SINC & 0.60 & \xdash & \xdash & \xdash & \xdash & 6e-01 & $\star$ & 0.0000000002 & \xdash & \xdash & 1e-02 & $\star$ & $\star$ & $\star$ & 0.0000000046\\
 COS  &      & \xdash & \xdash & \xdash & \xdash & \xdash & $\star$ & & \xdash &        \xdash & \xdash & \xdash & \xdash & \xdash & \\
\hline
 SINC & 0.70 & \xdash & \xdash & \xdash & \xdash & $\star$ & $\star$ & 0.0000000086 & \xdash & 3e-01 & 1e-04 & $\star$ & $\star$ & $\star$ & 0.0000002286 \\
 COS  &      & \xdash & \xdash & \xdash & \xdash & 1e-01 & $\star$ & & \xdash & \xdash & \xdash & \xdash & \xdash & 2e-01 & \\
\hline
 SINC & 0.80 & \xdash & \xdash & 8e-01 & 4e-03 & $\star$ & $\star$ & 0.0000005625 & \xdash & 8e-04 & 7e-06 & $\star$ & $\star$ & $\star$ & 0.0000188150 \\
 COS  &      & \xdash & \xdash & 5e-01 & 5e-02 & 2e-03 & $\star$ & & \xdash & \xdash & \xdash & \xdash & 9e-02 & 7e-04 & \\
\hline
 SINC & 0.90 & 7e-01 & 2e-01 & 2e-03 & 6e-04 & $\star$ & $\star$ & 0.0000422546 & 5e-02 & 1e-04 & 2e-07 & $\star$ & $\star$ & $\star$ & 0.0016079673 \\
 COS  &      & 4e-01 & 1e-01 & 1e-02 & 8e-04 & 1e-04 & $\star$ & & \xdash & 2e-01 & 1e-01 & 4e-02 & 7e-04 & 1e-04 & \\
\hline
 SINC & 1.00 & 4e-03 & 2e-03 & 1e-05 & 5e-06 & $\star$ & $\star$ & 0.0050767335 & 1e-03 & 4e-06 & 7e-09 & $\star$ & $\star$ & $\star$ & 0.3546252030 \\
 COS  &      & 4e-02 & 3e-03 & 1e-03 & 2e-04 & 6e-06 & 3e-07 & & 2e-01 & 4e-02 & 2e-03 & 1e-03 & 1e-04 & 4e-06 & \\
\hline
 SINC & 1.10 & 6e-05 & 1e-04 & 5e-07 & 3e-07 & 1e-09 & $\star$ & 0.1000001857 & 1e-04 & 3e-07 & 5e-10 & $\star$ & $\star$ & $\star$ & 0.9999913370 \\
 COS  &      & 5e-04 & 7e-05 & 8e-06 & 4e-07 & 7e-08 & 1e-09 & & 8e-03 & 8e-05 & 3e-04 & 1e-04 & 1e-06 & 3e-07 & \\
\hline
 SINC & 1.20 & 6e-04 & 9e-06 & 3e-06 & 1e-07 & $\star$ & $\star$ & 0.2000000119 & 1e-05 & 1e-07 & 1e-10 & $\star$ & $\star$ & $\star$ & 0.9999996919 \\
 COS  &      & 2e-04 & 2e-05 & 4e-07 & 5e-07 & 7e-09 & 2e-10 & & 2e-03 & 8e-04 & 2e-04 & 1e-05 & 3e-06 & 2e-07 & \\
\hline
 SINC & 1.30 & 4e-04 & 9e-06 & 2e-06 & 9e-08 & $\star$ & $\star$ & 0.3000000020 &  8e-06 & 1e-07 & 1e-10 & $\star$ & $\star$ & $\star$ & 0.9999999577 \\
 COS  &      & 4e-05 & 7e-06 & 1e-06 & 3e-07 & 5e-09 & 4e-10 & & 4e-03 & 5e-04 & 4e-05 & 1e-05 & 2e-06 & 1e-07 & \\
\hline
 SINC & 1.40 & 3e-04 & 8e-06 & 2e-06 & 7e-08 & $\star$ & $\star$ & 0.4000000005 & 8e-06 & 7e-08 & 6e-11 & $\star$ & $\star$ & $\star$ & 0.9999999902 \\
 COS  &      & 6e-05 & 4e-06 & 3e-07 & 2e-07 & 4e-09 & 3e-10 & & 2e-03 & 4e-04 & 2e-04 & 1e-05 & 2e-06 & 1e-07 & \\
\hline
\end{tabular}
\label{rHeston_conv_0.01}
\end{table}
   
\begin{figure}[h!]
	\centering
	\caption{PDF of the asset log-price under the rHeston model for $T=1$ (lhs) and $T=0.01$ (rhs). Parameters: $H=0.05,\nu=0.40,\rho=-0.65$.}
	\begin{subfigure}[b]{0.49\textwidth}
		\centering
		\includegraphics[width=\textwidth]{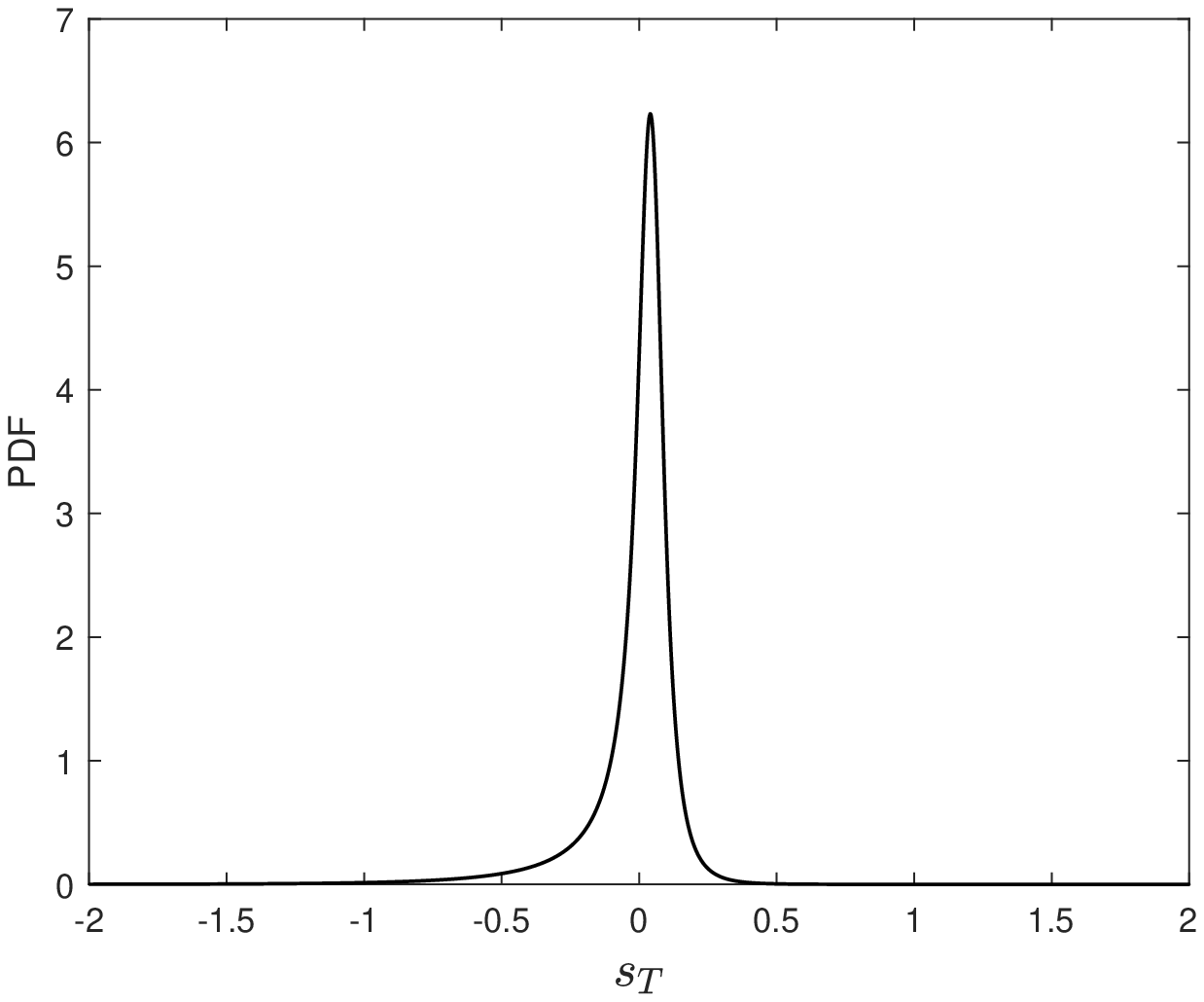}
	\end{subfigure}
	\begin{subfigure}[b]{0.49\textwidth}
		\centering
		\includegraphics[width=\textwidth]{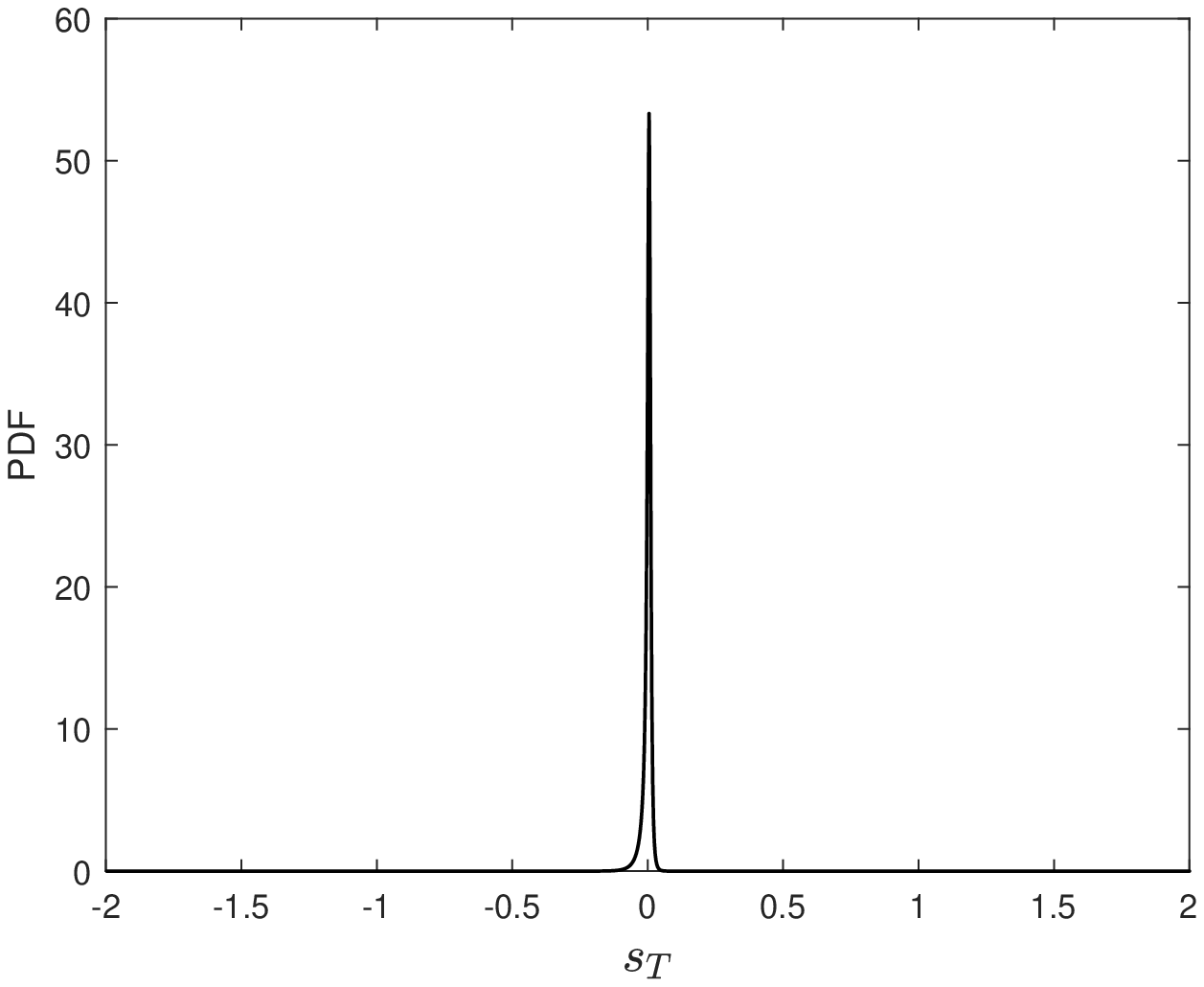}
	\end{subfigure}
	\label{pdf_rHeston}
\end{figure}

\noindent One difference with respect to the case $T=1$ that one can immediately spot is that digital options (and PV, as a consequence) tend to oscillate more with very short maturities. For example, when $\NCF=256$ neither \emph{SINC} nor COS provide meaningful values when the option is out-of-the-money. Again, this is a consequence of Fact 3 that we have listed. Figure (\ref{pdf_rHeston}) reports the PDFs for visual aid. \\

\noindent While this concludes our analysis for the convergence of the \emph{SINC} approach, one last comment is in order about Fact 2. We have seen that convenient selection of the bounds for the integration range guarantees (more or less) fast convergence to the "true" option price, and this is a strong empirical indication as not to worry about the contribution of error terms $\epsilon_2$ and $\epsilon_3$ (whose numerical characterization would critically depend on the particular model at hand). This very conforting practical regularity that we observe for all the models we have considered is sometimes violated by the rHeston model: when the expiration is long enough it is in fact true that the agreement between high precision \emph{SINC} and COS candidates is only ensured to reach 8/9 decimal digits. One way to fix this issue is to multiply the computed $X_c$ by some factor, which fact suffices to impute the cause of such a little lack of accuracy to $\epsilon_3$ (recall that the magnitude of $\epsilon_2$ only depends on $N$). Then, by the definition of $\epsilon_3$ we know that it associates with the replacement of the CF of the truncated density with the CF of the complete one, thus suggesting some little issue with the numerical approximation of the solution to the fractional Riccati equation which should be investigated more carefully.

\section{Accuracy and Efficiency of FFT-\emph{SINC}}\label{section:section 05}

If our earlier comparison with the COS method depicts the \emph{SINC} as a very accurate solution for Fourier pricing and (often times) it shows improved convergence with respect to the cosine expansion, one may be obviously interested in testing the FFT version of subsection \ref{FFT-SINC} against standard techniques of the same type. Availability of an FFT method is crucial for large scale problems such as calibration -- where several strikes have to be computed simultaneously for any given maturity -- and this is the main reason why Carr-Madan traditional technique or (a naive discretization of) Lewis formula are often preferred to the COS, for such practical purposes: at the end of the day, some lack of accuracy due to interpolation error may be an acceptable price to pay for the $N \log(N)$ computational complexity of the FFT algorithm. We therefore challenge FFT-\emph{SINC} againts such methodologies in the attempt to show that it requires much lower $\NCF$ to reach satisfactory accuracy on the implied volatilities. In case this is achieved, then we may well regard FFT-\emph{SINC} as a benchmark method in terms of efficiency as well. \\

\noindent Before we proceed in this direction we slightly elaborate on Lewis formula so as to clarify what we mean by a naive discretization. The reader will recognize Equation (\ref{Lewis_quadra}) as the price of a PV call option under \cite{lewisbook}:
\begin{align}
C(S,K,T) = S - \frac{\sqrt{SK}}{\pi} \int_0^{+\infty} \Re \bigg[ \e^{-iuk}\phi_T \bigg( u-\frac{i}{2} \bigg) \bigg] \frac{\dd u}{u^2+\frac{1}{4}}. \label{Lewis_quadra}
\end{align}
We are assuming zero interest rate and dividend yield here and denote $\phi_T(\cdot)$ the CF of the asset log-price. 
So, once a proper bound $u_{\max}$ has been defined for the integral above, and according to the strike grid
\begin{align*}
K_v=\e^{k_v}, \qquad k_v=-b+\gamma v \quad v=0,\dots,N-1 \quad b = \frac{N\gamma}{2} \quad \gamma = \frac{2\pi}{N\eta} \quad \eta = \frac{u_{\max}}{N-1} 
\end{align*} 
one can discretize Equation~(\ref{Lewis_quadra}) as 
\begin{align}
C(S,K_v,T) \approx S - \frac{\sqrt{SK_v}}{\pi} \sum_{j=0}^{N-1}{}^{'} \Re \bigg[ \e^{iu_jb} \e^{-i\frac{2\pi}{N}jv}\phi_T \bigg( u_j-\frac{i}{2} \bigg) \bigg] \frac{\eta}{u_j^2+\frac{1}{4}}, \qquad u_j=\eta j\label{eq:Lewis_discrete}
\end{align}
where $\sum{}^{'}$ indicates that the first term in the sum is multiplied by $\frac{1}{2}$. Simpson weights may be used as well (this is the suggestion from \cite{carr1999option}, for example), but we do not see any improvement when this is done. \\

\noindent This very much resembles the approach by \cite{carr1999option} in many respects and both are easily generalized to the frFFT if needed. The paper by~\cite{chourdakis2005option} provides the details for the implementation. \\

\noindent For the purposes of comparison, we now get back to the rough Heston model and price the same volatility surfaces as in \cite{el2019roughening}. We maintain the forward variance form that we have reported in the previous section and make the forward variance curve flat for simplicity\footnote{In a concrete situation the standard strategy is to estimate the forward variance curve as a difference on the variance swap curve. The fair value of a variance swap is computed using the methodologies explained in \cite{fukasawa2012normalizing} and an iteration procedure is subsequently performed to match model and market at-the-money volatilities through shifting and scaling.}.
We use suitable parameters in pricing, invert for the implied volatility and report the lowest $\NCF$ that we need to make the average absolute error on the volatilities of the order $10^{-6}$, over each smile. Then, the idea is that this will establish a hierarchy of the different methods. \\

\noindent As this will not appear in our study, benchmark prices are computed via high-precision COS; implied volatilities are calculated on those benchmarks and absolute differences referenced to them. \\

\noindent We have already noticed that the \emph{SINC} approach requires a suitable truncation range for the PDF of the asset log-price to be calculated: we do so by the cutting procedure that we have explained and repeat the same exercise for every maturity in the volatility surface. This has a precise meaning within the \emph{SINC} but it still may be (indirectly) linked to the choice of the upper limit of integration $u_{\max}$ with Carr-Madan method and Lewis formula. We focus on the latter, here. \\

\noindent The periodicity of the sum~(\ref{eq:Lewis_discrete}) approximating the Lewis integral may in fact be linked to the choice of $X_c$ for the \emph{SINC} and exploited to choose the step size in the discretization grid $\eta$. However, since Lewis formula does not leverage the bounded support of the PDF as the \emph{SINC} does, one should allow for an extra degree of freedom and accept the following definition
\begin{align*}
\eta = \frac{1}{2X_c\beta}
\end{align*}  
where $\beta$ is to be optimally chosen so as to minimize the average distance over the surface between Lewis price and the benchmark. We stress that the choice for $\beta$ depends on $\NCF$ and is to be repeated each time this is changed.\\ 

\noindent The exact same procedure holds true for Carr-Madan method but in general the optimal $\beta$'s will be different. Moreover, it is important to stress that the dumping parameter is also to be chosen for Carr-Madan. As a precise rule to fix it does not exist, we proceed by testing several values. The best performance we experience corresponds to the case $\alpha_{CM}=0.4$. \\

\noindent The relation between $\NCF$ and $u_{\max}$ (and $\alpha_{CM}$) is not clearly understood in Lewis and Carr-Madan method, and no theoretical or heuristic argument is available to properly select the upper limit of integration, making it difficult to blindly trust those approaches. In this regard, having a straightforward interpretation for $X_c$ in the \emph{SINC} and a direct link with the option price approximation error should also be very much appreciated.\\

\noindent Coming to our numerical experiments, we have two volatility surfaces to be priced. The first one is for the S\&P500 as of August 14, 2013 and it consists of 19 expirations from a couple of days to about 2.5 years, for a total 1291 strike-expiration pairs. Calibrated parameters in \cite{el2019roughening} are given as: 
\begin{align*}
H = 0.1216 \qquad \nu = 0.2910 \qquad \rho = -0.6714.
\end{align*}
With these numbers we compute put option prices for the entire surface based on all of FFT methods above. We also advocate the frFFT versions for a complete overview of the results, that we quote in Table \ref{FFT_20130814}. We report the lowest $\NCF$ to reach the required accuracy per smile described in the previous paragraph. \\

\noindent The computed $X_c$ vector for the \emph{SINC} is available on GitHub\footnote{https://github.com/fabioBaschetti/SINC-method. This also contains the surfaces that we use and all the codes one needs to reproduce our results.}, and the $\beta$'s that connect them to the upper limit of integration $u_{\max}$ in \cite{lewisbook} and \cite{carr1999option} methods are also reported in Table \ref{FFT_20130814}.

\begin{table}[h!]
\centering	
\caption{$\NCF$ values and $\beta$'s for average absolute errors on the implied volatilities to be at most $10^{-6}$ over each smile. \emph{SINC}, Lewis and Carr-Madan methods are investigated in both the FFT and frFFT versions. Fractional parameter $\epsilon$ is chosen $\epsilon= 0.15$ for \emph{SINC}, $\epsilon=0.02$ for the others. [S\&P500 index as of August 14, 2013 -- assumption: forward variance curve flat at 0.0320]}
\begin{tabular}{|c|c|c|c|c|c|c|}
\hline
& \multicolumn{3}{c|}{FFT} & \multicolumn{3}{c|}{frFFT} \\
\hline
& SINC & Lewis & Carr-Madan & SINC  & Lewis & Carr-Madan \\ 
\hline
$\NCF$ & 8192 & 65536 $\blacklozenge$ & $\bullet$ & 512 & 4096 & 16384 \\
\hline
$\beta$ &  & 1.600 & 4.000 &  & 2.200 & 5.500 \\
\hline
\end{tabular}
\label{FFT_20130814}
\end{table}

\noindent One immediately observes that FFT-\emph{SINC} meets the target with relatively small $\NCF$, and it is actually true that more precise volatilities can be easily calculated by just increasing it. On the other hand, Lewis formula and Carr-Madan method experience difficulties which are unknown to the \emph{SINC}. Specifically, a naive discretization of Lewis integral fails to guarantee the desired accuracy on the volatilities for very short maturities (as the black diamond indicates) and Carr-Madan method does not go further than an average error of the order $10^{-5}$, even at high $\NCF$ (this is seen in a bullet point in the table). Such issues are eventually solved resorting to the fractional FFT. Lewis formula now requires $\NCF=4096$ evaluations of the CF for each smile and Carr-Madan method also reaches the target. In any case the \emph{SINC} still proves to be largely superior, with $\NCF=512$ evaluations only. The fractional parameter ($\epsilon$) restricting the domain of the conjugate variable is made to be as small as one can, compatibly with the strike grid from the surface. The precise meaning of $\epsilon$ is more directly understood in Section \ref{appendix:fractional} of the Appendix. \\

\noindent Similar patterns are found when we move to May 19, 2017. This second surface counts 3352 strike-expiration pairs, which are distributed over 35 maturities to cover basically the same period as before. We now use the parameters from Section~\ref{rHeston_SINCvsCOS} 
\begin{align*}
H = 0.0500 \qquad \nu = 0.4000 \qquad \rho = -0.6500
\end{align*}
and also conserve a flat forward variance curve (at $\xi_0(t)=0.0256$).
While Table \ref{FFT_20170519} shows larger values of $\NCF$ for all methods, thereby spotting a more complicated volatility surface, the relations between \emph{SINC} and its competitors are unchanged. This corroborates our claims about the superior performance of the \emph{SINC}.  

\begin{table}[h!]
\caption{$\NCF$ values and $\beta$'s for average absolute errors on the implied volatilities to be at most $10^{-6}$ over each smile. \emph{SINC}, Lewis and Carr-Madan methods are investigated in both the FFT and frFFT versions. Fractional parameter $\epsilon$ is chosen $\epsilon= 0.15$ for \emph{SINC}, $\epsilon=0.02$ for the others. [S\&P500 index as of May 19, 2017 -- assumption: forward variance curve flat at 0.0256]}
\centering	
\begin{tabular}{|c|c|c|c|c|c|c|}
\hline
& \multicolumn{3}{c|}{FFT} & \multicolumn{3}{c|}{frFFT} \\
\hline
& SINC & Lewis & Carr-Madan & SINC  & Lewis & Carr-Madan \\ 
\hline
$\NCF$ & 16384 & 65536 $\blacklozenge$ & $\bullet$ & 2048 & 8192 & 16384 \\
\hline
$\beta$ &  & 1.000 & 2.500 &  & 1.170 & 3.168 \\
\hline
\end{tabular}
\label{FFT_20170519}
\end{table}


\section{Conclusions}\label{sec:concl}

The paper investigates the \emph{SINC} approach when pricing PV options. \emph{SINC} is shown to be superior to well-known benchmark methodologies. At variance with COS, it allows for an immediate extension to the FFT form. This is essential in any calibration exercise. Prompted by our results, we claim that \emph{SINC} is a promising approach, regarding both the precision it achieves and its numerical efficiency. The numbers we produce in Sections \ref{section:section 03} and \ref{section:section 05} leave little space for different interpretations, they cover enough models to support  the claim that the method is flexible enough to deal with jump-diffusion as well as rough Heston models, with the obvious alert (as it is for all Fourier-based techniques) that it can only be applied when the CF of the asset log-price is known either in analytic or semi-analytic form. \\

\noindent The idea behind \emph{SINC} is that one first writes put options as a linear combination of digital Asset-or-Nothing and Cash-or-Nothing options. The expectation defining their values is a convolution between the density of the asset log-return and the payoff function. Then, the convolution theorem for Fourier transforms guarantees that each price can be expressed as the integral over a shifted CF. By approximating the CF of the true density with the CF of a truncated PDF, one can fully exploit the potential of the Shannon Sampling Theorem.
It allows to represent the CF at any point by means of a discrete set of frequencies and express it as a Fourier-sinc expansion. The option price expressed in this form is the Modifed Hilber trasform of the sinc function that can be computed in close form yielding simple and compact formulas for digital and PV put option prices. Moreover, these formulas lend themselves to fast computation by means of FFT. The paper provides a rigorous proof of the convergence of the SINC formula to the correct option price when the support grows and the number of Fourier frequencies increases. It also investigates several technical prescriptions, such as the computation of truncation bounds by means of a cutting procedure based on the CDF. Conversely, if one wants to follow \cite{doi:10.1137/080718061} and their cumulants-based rule, we also provide a novel technique to compute them from the CF. The paper also addresses the issue of the sensitivity of the option prices to the number $\NCF$ of frequencies sampled in the Fourier-space.
Through an extensive pricing exercise, it assesses the superior performance of the \emph{SINC} approach with respect to the competitor COS methodology. As far as the FFT specification is concerned, the paper challenges \emph{SINC} against the FFT specification of the Lewis formula and the Carr-Madan approach. In both cases, \emph{SINC} proves to be accurate and robust to option's specification.

\newpage

\section*{Appendix}
\appendix

\section{Inverse Fourier Transform of the $\theta$ Function }\label{appendix-inv-fft-theta}

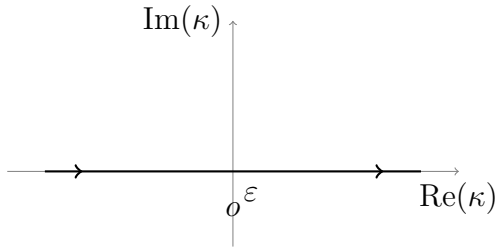
\begin{figure}[h!]
	\center
\caption{Possible integration path when integrand is $ \exp(i2\pi \kappa x )/(\kappa + i \varepsilon)$ \label{path_delta_m_alt}.}
	\begin{tikzpicture}[decoration={markings,
			mark=at position 0.5cm with {\arrow[line width=1pt]{>}},	
			mark=at position 4.5cm with {\arrow[line width=1pt]{>}}		
		}
	]

	\draw[help lines,->] (-3,0) -- (3,0) coordinate (xaxis);
	\node[below] at (xaxis) {$\operatorname{Re}(\kappa)$};
	\draw[help lines,->] (0,-1) -- (0,2) coordinate (yaxis);
	\node[left] at (yaxis) {$\operatorname{Im}(\kappa)$};

	\path[draw,line width=0.8pt,postaction=decorate] (-2.5, 0.0) -- (2.5,0);

	\node at (0.0, -.5) {$o$};
	\node[right] at (.0,-.3) {$\varepsilon$};

	\end{tikzpicture}
\end{figure}

Let us look at the distribution $\delta^-$ and let us recall the definition
\begin{equation*}
	\delta^-(\kappa) := \frac{i}{2\pi} \frac{1}{\kappa + i\varepsilon}.
\end{equation*}
In this appendix we want to show the following result:
\begin{equation} \label{ft_inv_delta_m}
 \theta(x)  = \int \dd \kappa\, \e^{-i2\pi \kappa x } \delta^-(\kappa).
\end{equation}
The term $i\varepsilon$ in the denominator of Equation~(\ref{ft_inv_delta_m}) is nothing but the prescription to follow
whenever the integration path runs over a singular point. The integral~(\ref{ft_inv_delta_m}) can be computed remaining on the real axis but moving the singularity on the negative imaginary axis as illustrated in Figure~(\ref{path_delta_m_alt}).
When $x < 0$  we can close the integration contour 
on the upper half plane as in Figure~(\ref{path_delta_m_x_neg}) and since there is no pole inside the
integration path the result is zero. On the other hand, when $ x > 0$ we can close the contour 
in the lower half plane as in Figure~(\ref{path_delta_m_x_pos}). Since we are running clockwise
the result will be:
\begin{eqnarray*}
	 \int \dd \kappa\, \e^{-i2\pi \kappa x } \delta^-(\kappa)  & = & \frac{i}{2\pi} \int_{\Gamma} \dd \kappa\, \e^{-i2\pi \kappa x }  \frac{1}{\kappa + i\varepsilon} \\
	                                        & = & \frac{i}{2\pi} \left[ -2\pi i \e^{-i 2\pi \kappa (-i\varepsilon)} \right] = 1 \quad x > 0.
\end{eqnarray*}

\begin{figure}[h!]
	\center
\caption{The integration path when integrand is $ \exp(-i2\pi \kappa x )/(\kappa + i \varepsilon)$  and $ x < 0$. \label{path_delta_m_x_neg}}
	\begin{tikzpicture}[decoration={markings,
			mark=at position 0.5cm with {\arrow[line width=1pt]{>}},	
			mark=at position 4.5cm with {\arrow[line width=1pt]{>}},	
			mark=at position 6.5cm with {\arrow[line width=1pt]{>}},
			mark=at position 11.5cm with {\arrow[line width=1pt]{>}}
		}
	]

	\draw[help lines,->] (-3,0) -- (3,0) coordinate (xaxis);
	\node[below] at (xaxis) {$\operatorname{Re}(\kappa)$};
	\draw[help lines,->] (0,-1) -- (0,3) coordinate (yaxis);
	\node[left] at (yaxis) {$\operatorname{Im}(\kappa)$};

	\path[draw,line width=0.8pt,postaction=decorate] (-2.5, 0.0) -- (2.5,0) arc ( 0:180:2.5);

	\node at (0.0, -.5) {$o$};
	\node[right] at (.1, -.3) {$\varepsilon$};

	\end{tikzpicture}
\end{figure}

\begin{figure}[h!]
	\center
\caption{The integration path when integrand is $ \exp(-i2\pi \kappa x )/(\kappa + i \varepsilon)$  and $x > 0$. \label{path_delta_m_x_pos}}
	\begin{tikzpicture}[decoration={markings,
			mark=at position 0.5cm with {\arrow[line width=1pt]{>}},	
			mark=at position 4.5cm with {\arrow[line width=1pt]{>}},	
			mark=at position 6.5cm with {\arrow[line width=1pt]{>}},
			mark=at position 11.5cm with {\arrow[line width=1pt]{>}}
		}
	]

	\draw[help lines,->] (-3,0) -- (3,0) coordinate (xaxis);
	\node[below] at (xaxis) {$\operatorname{Re}(\kappa)$};
	\draw[help lines,->] (0,-3) -- (0,1) coordinate (yaxis);
	\node[left] at (yaxis) {$\operatorname{Im}(\kappa)$};

	\path[draw,line width=0.8pt,postaction=decorate] (-2.5, 0.0) -- (2.5,0) arc ( 0:-180:2.5);

	\node at (0.0, -.5) {$o$};
	\node[right] at (.1, -.3) {$\varepsilon$};
	\node at (2.0, - 2.0) {$\Gamma$};

	\end{tikzpicture}
\end{figure}

\section{The Shannon Sampling Theorem}\label{appendix:Appendix A}

Let us consider a function $c(x)$ whose domain is centered around the origin, i.e. $c(x)$: $[-X_c,X_c] \to \mathbb{R}$. Its Fourier transform is defined as
\begin{align*}
\hat{c}(\kappa) = \int_{-X_c}^{X_c} \e^{i 2\pi \kappa x} c(x) \dd x,
\end{align*}
and the Fourier Inversion Theorem guarantees that the original function can be written 
\begin{align*}
c(x) = \frac{1}{2X_c} \sum_{n=-\infty}^\infty \hat{c}(\kappa_n)\e^{-i 2\pi \kappa_n x}.
\end{align*}
An immediate consequence is that
\begin{align*}
\hat{c}(\kappa) & = \frac{1}{2X_c} \sum_{n=-\infty}^\infty \hat{c}(\kappa_n) \int_{-X_c}^{X_c} \e^{i 2\pi (\kappa - \kappa_n) x} \dd x = \frac{1}{2X_c} \sum_{n=-\infty}^\infty \hat{c}(\kappa_n) \frac{\e^{i2\pi(\kappa - \kappa_n)X_c}-\e^{-i2\pi(\kappa - \kappa_n)X_c}}{i2\pi(\kappa - \kappa_n)}\\ & = \sum_{n=-\infty}^\infty \hat{c}(\kappa_n) \ \frac{\sin [2\pi (\kappa - \kappa_n) X_c]}{2\pi (\kappa - \kappa_n) X_c} \\
& = \sum_{n=-\infty}^\infty \hat{c}(\kappa_n) \text{sinc}[2\pi (\kappa - \kappa_n) X_c] = \sum_{n=-\infty}^\infty \hat{c}(\kappa_n) \text{sinc}[2\pi (\kappa_n - \kappa) X_c],
\end{align*}
where we have used that the sinc is an even function, in the last equality. \\
Similarly, for a function $z(x)$ defined over a bounded interval
\begin{align*}
I_z = \{x: X_l \leq x \leq X_h \},
\end{align*}
we get back to the same case as above by properly shifting the function $z$, i.e. 
\begin{align*}
c(x) \doteq z(x + X_m), \qquad X_m = \frac{X_h + X_l}{2}.
\end{align*}
Hence, knowledge of this next fact 
\begin{align*}
\hat{c}(\kappa_n) = \int_{-X_c}^{X_c} \e^{i2\pi \kappa_n x}c(x) \dd x = \e^{-i2\pi \kappa_n X_m} \int_{X_l}^{X_h} \e^{i2\pi \kappa_n x}c(x-X_m) \dd x = \e^{-i2\pi \kappa_n X_m}\hat{z}(\kappa_n) 
\end{align*}
makes it not difficult to show that 
\begin{align*}
\hat{z}(\kappa) & = \int_{X_l}^{X_h} \e^{i 2\pi \kappa x} z(x) \dd x = \int_{-X_c}^{X_c} \e^{i 2\pi \kappa (x+X_m)} z(x+X_m) \dd x \\
& = \e^{i 2\pi \kappa X_m} \int_{-X_c}^{X_c} \e^{i 2\pi \kappa x} c(x) \dd x = \e^{i 2\pi \kappa X_m} \hat{c}(\kappa) \\
& = \sum_{n=-\infty}^{\infty} \e^{i 2\pi \kappa X_m} \hat{c}(\kappa_n) \text{sinc}[2\pi(\kappa_n - \kappa)X_c] \\ 
& = \sum_{n=-\infty}^{\infty} \e^{i2\pi(\kappa - \kappa_n)X_m} \hat{z}(\kappa_n) \text{sinc}[2\pi(\kappa_n-\kappa)X_c],
\end{align*}

\section{The Modified Hilbert Transform}\label{appendix:Appendix B}

The object of our interest are integrals which take the following form
\begin{align*}
\int \frac{\text{sinc}[a(y-\kappa)]}{\kappa + i\varepsilon} \dd \kappa  
= \frac{2\pi}{i} \mathcal{H}^-[\text{sinc}(ay)], 
\end{align*}
that is the Modified Hilbert transform already introduced in Definition \ref{modHilb}.\\
Then
\begin{align*}
\mathcal{H}^-[\text{sinc}(ay)] & = \bar{\mathcal{F}} \big[ \mathcal{F}[\text{sinc}(ax)] \mathcal{F}[\delta^-(x)] \big]  \\
& = \frac{\pi}{|a|} \int \e^{-i 2\pi \kappa y} \theta(-\kappa) \mathbbm{1}_{[-\frac{|a|}{2\pi} < \kappa <\frac{|a|}{2\pi}]} \dd \kappa = \frac{\pi}{|a|} \int_{-\frac{|a|}{2\pi}}^0 \e^{-i 2\pi \kappa y} \dd \kappa \\ & = \frac{1}{-2i y |a|}(1-\e^{iy|a|}), 
\end{align*}
where we make use of the fact that the Fourier transform of the sinc function complies to
\begin{align*}
\mathcal{F}[\text{sinc}(ax)] & = \int \e^{i 2\pi \kappa x} \frac{\sin(ax)}{ax} \dd x = \int \e^{i 2\pi \kappa x} \frac{\sin(ax)}{ax+ai\varepsilon} \dd x \\ & = \frac{\pi}{a} \int \e^{i 2\pi \kappa x} \frac{\e^{iax}- \e^{-iax}}{2\pi i(x+i\varepsilon)} \dd x = \frac{\pi}{a} \bigg[ \theta \bigg( \kappa + \frac{a}{2\pi} \bigg) - \theta \bigg( \kappa - \frac{a}{2\pi} \bigg) \bigg] \\ 
& = \frac{\pi}{|a|} \mathbbm{1}_{[-\frac{|a|}{2\pi} < \kappa <\frac{|a|}{2\pi}]}.
\end{align*}
This is indeed justified by the fact that the sinc function is regular in the origin, so that we can shift the pole everywhere we want on the imaginary axis (and change the contour accordingly) without affecting the integral. \\
We consequently conclude that our target integral admits solutions of an exponential type
\begin{align*}
\int \frac{\text{sinc}[a(y-\kappa)]}{\kappa + i\varepsilon} \dd \kappa 
= \frac{\pi}{y|a|}(1-\e^{iy|a|}).
\end{align*}
Choosing $a = 2\pi X_c$ and $y=\kappa_n$ finally proves the desired result of Equation~(\ref{sinc_Hilbert}). 

\section{An Explicit Formulation for the CoN Put Price}\label{appendix:Appendix C}

This section derives an explicit formulation of the CoN put price, in terms of sin and cos functions multiplying real and imaginary parts of the Fourier transform $\hat{f}$. So, if we denote $\hat{f}^\dag(\cdot)$ the complex conjugate of the Fourier transform $\hat{f}(\cdot)$, we immediately have 
\begin{align*}
\mathbb{E}[\mathbbm{1}_{\{s_T<k\}}] & \simeq \frac{i}{2\pi} \sum_{n=-N/2}^{N/2} \e^{-i2\pi k \kappa_n} \hat{f}(\kappa_n) \bigg[ -i\pi \mathbbm{1}_{n=0} + \frac{1-(-1)^n}{n} \mathbbm{1}_{n\neq 0} \bigg] \\
& = \frac{1}{2} + \frac{i}{2\pi} \sum_{n=1}^{N/2} \frac{(1-(-1)^n)}{n} \bigg[ \e^{-i2\pi k \kappa_n} \hat{f}(\kappa_n) - \e^{i2\pi k \kappa_n} \hat{f}^\dagger(\kappa_n) \bigg]\,,
\end{align*}
which can be rewritten as
\begin{equation*}
\hspace{28mm} = \frac{1}{2} + \frac{i}{\pi} \sum_{n=1}^{N/4} \frac{1}{2n-1} \bigg[ \e^{-i2\pi k \kappa_{2n-1}} \hat{f}(\kappa_{2n-1}) - \e^{i2\pi k \kappa_{2n-1}} \hat{f}^\dagger(\kappa_{2n-1}) \bigg]\,.
\end{equation*}
Properly rearranging terms based on Euler's formula, we obtain
\begin{align*}
\mathbb{E}[\mathbbm{1}_{\{s_T<k\}}] \simeq \frac{1}{2} - \frac{2}{\pi} \sum_{n=1}^{N/4} \frac{1}{2n-1} \bigg[ & \cos(2\pi k \kappa_{2n-1}) \Im \big[\hat{f}(\kappa_{2n-1})\big] \ + \\ 
& - \sin(2\pi k \kappa_{2n-1}) \Re\big[\hat{f}(\kappa_{2n-1})\big] \bigg]. 
\end{align*}

\section{Numerical Moments of $q$-th Order}\label{appendix:Appendix D}

The computation of the moments of a distribution requires to manage integrals which are not always ensured to admit a closed form solution. Nevertheless, the knowledge of the CF allows to evaluate them numerically. This fact is of crucial importance if one wants to truncate the PDF according to the cumulants-based rule of \cite{doi:10.1137/080718061} but should be clearly recognized to have a much wider scope. That is why we suppress dependence on $s_T$ and talk about a random variable $X$ defined over the support $[-X_c,X_c]$, in this section. \\

\noindent Let us first recall the next fundamental relation between the $q$-th order moment of $X$ and its CF $\phi_X$:
\begin{align} \label{q-th_moment}
\mathbb{E}[X^q] & = (i2\pi)^{-q} \dvq{q}{\kappa} \phi_X(\kappa) \bigg\rvert_{\kappa=0} \nonumber \\
& \hspace{-40mm} \text{then, if we apply the Shannon Sampling Theorem} \nonumber \\
& = (i2\pi)^{-q} \sum_{n=-\infty}^{\infty} \phi_X(\kappa_n) \dvq{q}{\kappa} \text{sinc}(2\pi(\kappa_n-\kappa)X_c) \bigg\rvert_{\kappa=0} \nonumber \\
& \hspace{-40mm} \text{and perform a simple change of variable, we have} \nonumber \\
& = (iX_c)^q \sum_{n=-\infty}^{\infty} \phi_X(\kappa_n) \dvq{q}{t} \text{sinc}(t) \bigg\rvert_{t=n\pi}. 
\end{align}  
Furthermore, a power series expansion of the sinc function, i.e. 
\begin{align*}
\text{sinc}(t) = \sum_{n=0}^\infty (-1)^n \frac{t^{2n}}{(2n+1)!}
\end{align*}
is readily obtained given the corresponding expansion for the sin function, and this clearly justifies a number of properties. Among them we have the following: 
\begin{tasks}[counter-format={$\bullet$}, label-align=left, label-offset={-5mm}, label-width={12mm}, item-indent={10mm}, label-format={\bfseries}, column-sep=0mm](2)
\task odd derivatives are such that \begin{align*} \dvq{2q+1}{t} \text{sinc}(t) \bigg\rvert_{t=0} = 0 
\end{align*} by parity of the sinc function
\task terms of the following type  \begin{align*} \dvq{2q+1}{t} \text{sinc}(t) \bigg\rvert_{t=n\pi} 
\end{align*} are odd with respect to $n$
\task even derivatives are such that \begin{align*} \dvq{2q}{t} \text{sinc}(t) \bigg\rvert_{t=0} = \frac{(-1)^q}{2q+1}
\end{align*} by the theory of Taylor series
\task  terms of the following type \begin{align*} \dvq{2q}{t} \text{sinc}(t) \bigg\rvert_{t=n\pi}
\end{align*} are even with respect to $n$
\end{tasks}
These properties play a fundamental role when specifying Equation~(\ref{q-th_moment}) for some given $q$.
We report the explicit formulation of the first few moments next:

\begin{eqnarray*}
m_1 &= &\mathbb{E}[X] = -2X_c \sum_{n=1}^{\infty} \Im \big[ \phi_X(\kappa_n) \big] \frac{(-1)^n}{n\pi}\,,\\
m_2 &= &\mathbb{E}[X^2] = \frac{X_c^2}{3} + 4X_c^2 \sum_{n=1}^{\infty} \Re \big[ \phi_X(\kappa_n) \big] \frac{(-1)^n}{(n\pi)^2}\,,\\
m_3 &=& \mathbb{E}[X^3] = -2 X_c^3 \sum_{n=1}^{\infty} \Im \big[ \phi_X(\kappa_n) \big] \bigg[ \frac{(-1)^n}{n\pi} \bigg( 1-\frac{6}{(n\pi)^2} \bigg) \bigg]\,,\\
m_4 &= &\mathbb{E}[X^4] = \frac{X_c^4}{5} + 8X_c^4 \sum_{n=1}^{\infty} \Re \big[ \phi_X(\kappa_n) \big] \bigg[ \frac{(-1)^n}{(n\pi)^2} \bigg( 1-\frac{6}{(n\pi)^2} \bigg) \bigg]\,.
\end{eqnarray*}

\section{Error Analysis (proof)}\label{appendix:error}

The overall error $\epsilon$ is equal to the sum $\epsilon_1+\epsilon_2+\epsilon_3$ and its norm can be bounded as 
$$|\epsilon|\leq \epsilon_1 + |\epsilon_2| + |\epsilon_3|\,.$$ 
Arguing in the same way as in the COS paper, $\epsilon_1$ can be made arbitrarily small by choosing a sufficiently high value for $X_c$. 
As far as $\epsilon_2$ is concerned, it is clear from Equation~(\ref{CoN_02}) that it corresponds to the remainder of a series converging to $\mathbb{E}[\mathbbm{1}_{\{s_T<k\}} \mathbbm{1}_{\{-X_c\leq s_T \leq X_c\}}]$. Then, when $N$ increases, $\epsilon_2$ goes to zero~\footnote{
It is possible to derive an analytic bound for $\epsilon_2$, assuming some mild regularity for the PDF. The reasoning is similar to that in~\cite{doi:10.1137/080718061}.}.\\ 

Concerning $\epsilon_3$, one has
\begin{eqnarray*}
    |\epsilon_3| &\leq&  \frac{1}{\pi} \sum_{n =-N/4}^{N/4} \frac{1}{|2n-1|} \left|\reallywidehat{f\mathbbm{1}_{\{-X_c\leq s_T \leq X_c\}}}(\kappa_{2n-1}) - \hat{f}(\kappa_{2n-1})\right|\,.
\end{eqnarray*}
To bound the last quantity, we can proceed following two strategies, which are based upon different assumptions. 
We first recall that  
$$
\hat{f}(\kappa_{2n-1})-\reallywidehat{f\mathbbm{1}_{\{-X_c\leq s_T \leq X_c\}}}(\kappa_{2n-1})=\int_{\mathbb{R}\backslash [-X_c,X_c]} f(s_T) \e^{i2\pi \kappa_{2n-1} s_T} \dd s_T\,.
$$
To ensure convergence of AoN and PV call prices, for $s_T>>1$ the PDF $f(s_T)$ has to satisfy
$$
f(s_T)\leq C \e^{-\beta s_T}\,,
$$ 
with $C> 0$ and $\beta>1$. For $s_T<<-1$, we assume the following condition -- typically satisfied by commonly used stochastic models for log-returns 
$$
f(s_T) \leq C \e^{\gamma s_T}\,,
$$
with $\gamma>0$. 
Then,
\begin{align*}
|\epsilon_3|&\leq \frac{1}{\pi} \sum_{n =-N/4}^{N/4} \frac{1}{|2n-1|} \left|\int_{\mathbb{R}\backslash [-X_c,X_c]} f(s_T) \e^{i2\pi \kappa_{2n-1} s_T} \dd s_T\right|\\ 
&\leq \frac{1}{\pi} \sum_{n =-N/4}^{N/4} \frac{1}{|2n-1|} \int_{\mathbb{R}\backslash [-X_c,X_c]} f(s_T) \dd s_T \leq \frac{2}{\pi} \sum_{n = 0}^{N/4} \frac{1}{2n+1} \int_{\mathbb{R}\backslash [-X_c,X_c]} f(s_T) \dd s_T \\ 
& \leq \frac{1}{\pi} (2+\log(N/2+1)) \int_{\mathbb{R}\backslash [-X_c,X_c]} f(s_T) \dd s_T\\
& \leq \frac{C}{\pi} (2+\log(N/2+1)) \left(\frac{1}{\gamma}\e^{-\gamma X_c}+\frac{1}{\beta}\e^{-\beta X_c}\right)\,.
\end{align*}
Naming $\delta=\min(\beta, \gamma)>0$, we obtain
$$
|\epsilon_3|\leq \frac{C}{\pi} (2+\log(N/2+1)) \e^{-\delta X_c}\,.
$$
To conclude, it is sufficient to choose $X_c$ proportional to $\log (N/2+1)$. Practically, this assumption amounts to choose $L$ proportional to $\log (N/2+1)$ in~(\ref{trunc_rule}). Then, $\epsilon_3$ can be made arbitrarily small by increasing $N$.\\
An alternative strategy allows to reach the same conclusion, without assuming the dependence of $X_c$ on $N$, but under a different hypothesis about the asymptotic behavior of the density $f(s_T)$. We can split the integral $\int_{\mathbb{R}\backslash [-X_c,X_c]} f(s_T) \e^{i2\pi \kappa_{2n-1} s_T} \dd s_T$ in two terms, $I_1$ and $I_2$, with
$$I_1(\kappa_{2n-1}) = \int_{-\infty}^{-X_c} f(s_T) \e^{i2\pi \kappa_{2n-1} s_T} \dd s_T\,\quad\text{and}\quad I_2(\kappa_{2n-1}) = \int_{X_c}^{+\infty} f(s_T) \e^{i2\pi \kappa_{2n-1} s_T} \dd s_T\,,$$
so that 
\begin{equation}\label{eq:epsilon_3}
|\epsilon_3| \leq \frac{1}{\pi} \sum_{n =1}^{N/4} \frac{1}{2n-1} \left|I_1(\kappa_{2n-1}) + I_2(\kappa_{2n-1})\right|+
\frac{1}{\pi} \sum_{n =1}^{N/4} \frac{1}{2n+1} \left|I_1^\dag(\kappa_{2n+1}) + I_2^\dag(\kappa_{2n+1})\right|\,.
\end{equation}
Let us consider $I_2(\kappa_{2n-1})$ and define the variable $y$ via the relation
$$s_T = y + \frac{X_c}{2n-1}\,.$$
Then,
\begin{eqnarray}
   I_2(\kappa_{2n-1}) & = & -\int_{X_c - X_c/(2n-1)}^{+\infty} \e^{i2\pi \kappa_{2n-1} y} f\left(y + \frac{X_c}{2n-1} \right)\, \dd y \nonumber
   \\  & = & -\int_{X_c}^{+\infty} \e^{i2\pi \kappa_{2n-1} y} f\left(y + \frac{X_c}{2n-1} \right)\, \dd y -\int_{X_c-X_c/(2n-1)}^{X_c} \e^{i2\pi \kappa_{2n-1} y} f\left(y + \frac{X_c}{2n-1} \right)\, \dd y\,. \nonumber
\end{eqnarray}
It follows that
\begin{eqnarray*}
   I_2(\kappa_{2n-1}) &=& \frac{1}{2}\int_{X_c}^{+\infty} \e^{i2\pi \kappa_{2n-1} y} \left(f\left(y\right)-f\left(y + \frac{X_c}{2n-1} \right)\right)\, \dd y\\
    &&-\frac{1}{2}\int_{X_c-X_c/(2n-1)}^{X_c} \e^{i2\pi \kappa_{2n-1} y} f\left(y + \frac{X_c}{2n-1} \right)\, \dd y\,\,
\end{eqnarray*}
so 
\begin{equation}
   |I_2(\kappa_{2n-1})|\leq\frac{1}{2}\int_{X_c}^{+\infty} \left|f\left(y\right)-f\left(y + \frac{X_c}{2n-1} \right)\right|\,\dd y +\frac{1}{2}\int_{X_c-X_c/(2n-1)}^{X_c} f\left(y + \frac{X_c}{2n-1} \right)\,\dd y\,. \nonumber
\end{equation}
We now assume that $f(s_T)$ is monotonically convergent to zero for sufficiently large $|s_T|$. The argument of the modulus is positive, so
\begin{eqnarray}
   |I_2(\kappa_{2n-1})|  & \le &  \frac{1}{2}\int_{X_c}^{X_c\left(1+\frac{1}{2n-1}\right)} f(s_T) \,\dd s_T
          + \frac{1}{2}\int_{X_c}^{X_c\left(1+\frac{1}{2n-1}\right)} f(s_T) \,\dd s_T\le  \frac{X_c}{2n-1}f(X_c)\,.\nonumber
\end{eqnarray}
Defining $s_T = y - X_c/(2n-1)$, it readily follows that 
$$|I_1(\kappa_{2n-1})|\leq \frac{X_c}{2n-1}f(-X_c)\,.$$ 
Similar results hold for $I_1^\dag(\kappa_{2n+1})$ and $I_2^\dag(\kappa_{2n+1})$. \\
From Equation~(\ref{eq:epsilon_3}), we obtain
\begin{eqnarray*}
|\epsilon_3| &\leq & \frac{X_c}{\pi}\left(f(X_c)+f(-X_c)\right) \sum_{n=1}^{N/4}\left(\frac{1}{(2n-1)^2}+\frac{1}{(2n+1)^2}\right)\\
\end{eqnarray*}
and, based on the following observation\footnote{We thank an anonymous referee for pointing this out.}
\begin{align*}
\sum_{n=1}^{\infty}\left(\frac{1}{(2n-1)^2}+\frac{1}{(2n+1)^2}\right) = \frac{\pi^2}{4}-1
\end{align*} 
we write \\
$$
|\epsilon_3| \leq  \frac{X_c}{\pi}\left(f(X_c)+f(-X_c)\right) \Upsilon,
$$
where $\Upsilon$ is a moderate constant (certainly smaller than $\pi^2 / 4 -1$). \\

To conclude, it is sufficient to assume  the existence of the first moment of $s_T$. Indeed, this implies that $f(s_T) = o(1/s_T)$ for $|s_T| \rightarrow +\infty$. Then, $X_c f(X_c)$ and $X_c f(-X_c)$  can be made arbitrarily small by choosing $X_c$ sufficiently large.

\section{The Fractional Fourier Transform}\label{appendix:fractional}

Throughout this paper we had to compute double infinite sums of the type:
\begin{equation*}\label{fourier-series}
	p(x) = \frac{1}{2X_c}\sum_{n=-\infty}^{+\infty} \hat{p}_n e^{- i2\pi k_n x} \quad 0 \le x \le 2X_c,\quad\text{with}\quad k_n = \frac{n}{2 X_c}
\end{equation*}
The $p_N$ approximation to $p(x) $ is given by:
\begin{equation*}\label{N-term-fourier-series}
	p_N(x) = \frac{1}{2X_c}\sum_{n=-N/2}^{N/2} \hat{p}_n e^{-i2\pi k_n x}.
\end{equation*}
If we confine our interest to the discrete set of values:
\[
	x_m = m \frac{2X_c}{N} , \quad -N/2 \le m < N/2 
\]
we get:
\begin{eqnarray*}
	p_N(x_m) & = &  \frac{1}{2X_c}\sum_{n=0}^{N-1} \hat{q}_n e^{-i2\pi n m/N}  \label{discrete-fft-form}
    \\ q_n & = & \left\{    \begin{array}{lr} 
                                              p_n & 0 \le n < N/2,
                                             \\ p_{N/2} + p_{-N/2} & n = N/2,
                                              \\ p_{n-N} & N/2 < n < N.
                            \end{array}
                 \right. \nonumber
\end{eqnarray*}
and this sum is what we compute with the FFT.

\noindent The fractional Fourier transform computes:
\begin{equation*}\label{ft_fract}
	p_N(\hat{x}_m) = \frac{1}{2X_c} \sum_{n=-N/2}^{N/2-1} \hat{p}_n e^{-i2\pi n m \epsilon/N}.
\end{equation*}
where $\hat{x}_m = m\epsilon \delta x $.

\end{document}